\documentclass[12pt,preprint]{aastex}

\shorttitle{SMC stellar populations}
\shortauthors{No\"el, N.E.D. et al.}

\begin{document}


\title{OLD MAIN-SEQUENCE TURNOFF PHOTOMETRY IN THE SMALL MAGELLANIC CLOUD. I. 
CONSTRAINTS ON THE STAR FORMATION HISTORY IN DIFFERENT FIELDS}

\author{Noelia E. D. No\"el}
\affil{Instituto de Astrof\'\i sica de Canarias. 38200 La
Laguna. Tenerife, Canary Islands. Spain.}
\email{noelia@iac.es}

\author{Carme Gallart} 
\affil{Instituto de Astrof\'\i sica de Canarias. 38200 La
Laguna. Tenerife, Canary Islands. Spain.}
\email{carme@iac.es}

\author{Edgardo Costa}
\affil{Departamento de Astronom\'\i a, Universidad de Chile, Casilla 36-D, Santiago, Chile.}
\email{costa@das.uchile.cl}

\and

\author{Ren\'e A. M\'endez}
\affil{Departamento de Astronom\'\i a, Universidad de Chile, Casilla 36-D, Santiago, Chile.}
\email{rmendez@das.uchile.cl}



\begin{abstract}

We present ground-based {\it B} and {\it R}-band color-magnitude diagrams (CMDs),
 reaching the oldest main-sequence turnoffs with good photometric accuracy for twelve fields in the Small Magellanic Cloud (SMC). Our fields, located between $\sim$1${\degr}$ and $\sim$4${\degr}$ from the center of the galaxy, are situated in different parts of the SMC such as the ``Wing'' area, and towards the West and South.  In this paper we perform a first analysis of the stellar content in our SMC fields through 
comparison with theoretical isochrones and color functions (CFs). We find that the underlying spheroidally distributed 
population is composed of both intermediate-age and old stars and that its age composition does not show strong galacto- 
centric gradients. The three fields situated toward the east, in the Wing region, show very active current star formation. However, only in the eastern field closest to the center do we find an enhancement of recent star formation with respect to a 
constant SFR(t). The fields corresponding to the western side of the SMC present a much less populated young MS, and 
the CF analysis indicates that the SFR$(t)$ greatly diminished around 2 Gyr ago in these parts. Field smc0057, the closest 
to the center of the galaxy and located in the southern part, shows recent star formation, while the rest of the southern 
fields present few bright MS stars. The structure of the red clump in all the CMDs is consistent with the large amount 
of intermediate-age stars inferred from the CMDs and color functions. None of the SMC fields presented here are 
dominated by old stellar populations, a fact that is in agreement with the lack of a conspicuous horizontal branch in all 
these SMC CMDs. This could indicate that a disk population is ruling over a possible old halo in all the observed fields.

\end{abstract}


\keywords{local group galaxies: evolution --- galaxies: individual (Small Magellanic Cloud) --- galaxies: photometry --- galaxies: stellar content --- Local Group}


\section{INTRODUCTION} \label{intro}

Dwarf galaxies are believed to represent the dominant population, by
number, of the present day universe, and a major constituent of groups
(C\^ot\'e et al. 1997) and clusters (e.g. Ferguson \& Sandage 1991).  Studying their star formation and chemical enrichment histories is key to understanding the evolution of galaxies on cosmological timescales (e.g. Madau et al. 1998).  Local Group galaxies are ideal laboratories for detailed studies of 
dwarf galaxy properties: we can resolve their individual stars and 
thus learn about their star formation histories (SFHs) by exploring ages, metallicities, and the spatial distribution of the stellar populations they contain. 

The color-magnitude diagram (CMD) is the best tool to retrieve the SFH
of a stellar system. The ideal situation is when it reaches the oldest
main-sequence (MS) turnoffs with good photometric accuracy.  In this
case, the information on the SFH can be obtained directly and with
little ambiguity from the distribution of stars on the MS, and its
comparison with that predicted by stellar evolution models in this
relatively well known phase of stellar evolution. In particular, the
range of ages and metallicities present can be determined through
comparison with theoretical isochrones. To quantitatively
determine the SFH, it is necessary to compare the observed density
distribution of stars with that predicted by stellar evolution models
(see Gallart, Zoccali \& Aparicio 2005). Shallower CMDs (e.g. those
reaching just below the horizontal-branch (HB) or a couple of
magnitudes below the tip of the red giant branch (RGB)) still contain
stars born through the whole galaxy's lifetime, but the interpretation
of the distribution of stars in terms of the SFH is progressively less
detailed and uncertain as the CMD is shallower. The reason for this is
twofold: on one hand, the stellar evolution models are less accurate
for more advanced stellar evolution phases such as the RGB or the HB,
because the corresponding physics is more complicated or uncertain. On
the other hand, in these stellar evolution phases, stars of very
different ages are packed together in the CMD in a small interval of
color and/or magnitude, and suffer from important age-metallicity
degeneracies (see Gallart 2000 and Gallart et al. 2005 for
detailed discussions of all these issues).

For all the Milky Way satellites, it is possible to obtain CMDs
reaching the oldest MS turnoffs using ground based telescopes, while
for the rest of the Local Group, this is possible using the ACS on
board the HST. The Magellanic Clouds (MCs) are particularly interesting
among the Milky Way satellites since they are actively star forming
galaxies which SFH can shed light on the role played by interactions
in galaxy evolution. But in spite of their proximity and intrinsic
interest, it is remarkable that there are still important gaps in the
knowledge of the SFH of the MCs.  This may be explained by their huge
projected size and the big number of stars to analyze.

In this paper we will focus on the Small Magellanic Cloud (SMC). The SMC is a dwarf irregular satellite of our Galaxy,
 with low mean metallicity and a high mass fraction remaining in gaseous form (van den Bergh 1999).
 These characteristics suggest that the SMC is in a more primitive evolutionary stage than its larger neighbor, 
 the Large Magellanic Cloud (LMC).
  Historically, the study of this galaxy has been in general
 neglected in favor of the LMC. The uncertainty about its 
 line of sight depth, its more complicated shape (elongated towards the Northeast), and its larger distance from us are factors at
 play. 
 
 To our knowledge, only two papers have presented CMDs reaching the oldest
MS turnoffs for a small field of view each: Dolphin et al. (2001) and
McCumber et al. (2005). Dolphin et al. presented a
combination of HST and ground-based {\it V} and {\it I} images of a SMC field 
situated 2${\degr}$ Northeast of NGC 121 (avoiding the contamination
from 47 Tucanae) in order to derive the SFH of this SMC field. From their ground-based images,
they inferred a peak of star formation between 5 and 8 Gyr ago with a medium age of 7.5 Gyr at the 1 $\sigma$ level,
 and that 14$\pm$5\% of the stars were formed more
than 11 Gyr ago. The SFH they derived is also consistent at the 2 $\sigma$ level with 
a continuous star formation from ancient times until $\sim$2 Gyr ago, when the star formation rate
 dropped to finally stop entirely in the past 0.5 Gyr. 
  More recently, 
 McCumber et al. (2005) analyzed the stellar populations of a SMC field located in the ``Wing''\footnote{We call ``Wing''
 area to the Eastern part of the SMC. Some authors (in particular, Gardiner \& Hatzidimitriou (1992) and papers in the same series) 
 also distinguish between the inner and outer ``Wing'' and the ``Arm''.}
  area with observations from the 
 HST WFPC2. They compared the luminosity function from 
 their observed CMD with those obtained from two different model CMDs, one with constant SFR(t)
  and another with bursts of star formation  
 at $\sim$2 and at $\sim$8 Gyr.
They found that the population appears to have formed largely in a
 quasi-continuous mode, with a main period of star formation between 4-12 Gyr ago and very recent star formation with 
 bright stars as young as 100$\pm$10 Myr.

Two main wide field studies of the SMC exist: the pioneer work 
by Gardiner \& Hatzidimitriou (1992) and a recent study from Harris \& Zaritsky (2004; hereafter HZ04). 
 Gardiner \& Hatzidimitriou (1992) presented the first large area study of the SMC, and mainly concentrated their analysis
  in the outer regions,
  beyond 2${\degr}$ from the SMC center.
  They took six photographic plates with the UK Schmidt Telescope covering a total area of 130${\degr}$$\times$130${\degr}$.
 Since their photometry is rather shallow (reaching the HB level at R$\sim$20 mag), they mainly gave information 
about the young populations (age $\leq$2 Gyr). From their CMDs 
 and contour plots of the surface distribution of MS stars with {\it B-R}$<$0.1 and {\it R}$<$20, they noticed the almost complete
 absence of bright MS stars in the Northwestern part, while a considerable bright 
 MS population was present in the Eastern and Southern area. 
 With the aid of luminosity functions they found that young populations ($<$0.6 Gyr in age) are concentrated towards the center of the SMC
 and in the ``Wing'' region.
Using an index defined as the difference between the 
median color (in {\it (B-R)}) of the RC and the color of the RGB 
at the level of the HB, the authors inferred that the bulk of the field population 
  has a median age around 10-12 Gyr. 
 They were the first to notice the different distribution of
  young and old populations in the SMC. Zaritsky et al. (2000) and Cioni et al. (2000) confirmed that the asymmetric appearance 
 of the SMC, similar to that 
of the HI (Stanimirovi\'c et al. 1999), is primarily caused by the
 distribution of young stars (upper MS stars, younger than $\sim$0.2 Gyr),
 and that the older stars have a spheroidal distribution. 
 
HZ04 presented a study of the SMC based on the Magellanic Clouds Photometric Survey (MCPS)
 {\it UBVI} catalog (Zaritsky et al. 1997), which covers a 4${\degr}$$\times$4.5${\degr}$ area of the SMC to a depth of {\it V}$\leq$21. 
This work represents the most complete analysis of the spatially resolved SFH of the central part of the SMC. 
 They divided the SMC survey in a grid of 351 cells and obtained the SFH, through a $\chi$$^2$ minimization 
between the star counts in the observed CMD and those in the model CMDs based on isochrones by Girardi et al. (2002).  
 They inferred that a significant fraction ($\sim$50\%) of all stars in the SMC were formed more than
  8.4 Gyr ago. Between 3 and 8.4 Gyr ago they found a quiescent period during which the SMC formed few stars. 
They also found a recent active time from 3 Gyr ago until now, with bursts at 2.5 Gyr, 400 Myr, and 60 Myr. 
 Although deeper than the one from Gardiner \& Hatzidimitriou (1992), this study is still limited by its shallowness.
 In fact, the oldest stars that they can observe on the MS are $\sim$2 Gyr old, and this implies a necessarily poor temporal
 resolution and larger uncertainties in their SFH at ages older than that. 
 
 In spite of the increasing interest in studying the SMC, reflected by the recent works
 mentioned before, there are still many questions unanswered which require the information provided by
  CMDs reaching the oldest MS turnoffs. 
 What is the age distribution of the old and intermediate-age population?, are there gradients in the composition of
 this underlying population?.
 Shallower studies
 inform about the existence of a large amount of young population in certain areas,
  but does this young population reflect an exceptional increase of the
 star formation at the present time with respect to the average SFR?.
   To shed light on these aspects, in this article we present twelve unprecedented deep
{\it BR}-based SMC CMDs with positions 
ranging from $\sim$1{\degr} to $\sim$4{\degr} from the SMC center.
 Our fields are distributed in different parts of the SMC, avoiding the central area,
 where crowding may not allow us to reach the oldest MS turnoffs from the ground. 
   Three of our fields are located in the ``Wing'' area, near the field studied by McCumber et al. (2005);  
 two fields are in the Western part and one in the Northwestern area, near the field from Dolphin et al. (2001) located 2{\degr} northeast
 of NGC121;
finally, six fields are placed towards the South at different galactocentric distances.
  This strategic selection of fields, in spite of their small size, permit us to sample several
  regions of the SMC with high temporal resolution in order to address the above issues. 
    The depth of our fields, reaching the oldest MS turnoffs with very
good photometric accuracy, will allow us to obtain detailed SFHs from all of them, and to investigate, therefore, the variation of the
SFH across the SMC field. This will be the subject of a forthcoming paper. Here we
 will focus the analysis in the inspection of the CMDs with the aid of theoretical isochrones and color functions. 
 
 This paper is organized as follows. In \S~\ref{obs} we present the observations and data
reduction. In \S~\ref{photometry} we discuss the photometry of the SMC fields, obtained using DAOPHOT II/ALLSTAR
and ALLFRAME programs. In \S~\ref{stellar} the CMDs are discussed with the help of isochrones and color functions.
  In \S~\ref{summary} and \S~\ref{conclu} our results are summarized and discussed.
 In a future paper (No\"el et al., in preparation) we will derive the SFH of the SMC through comparison of the
observed CMDs with synthetic CMDs.

\section{OBSERVATIONS AND DATA REDUCTION} \label{obs}

The present paper is a result of a more comprehensive program to study the SMC, which also includes the determination of its Absolute
Proper Motion by observing SMC fields in which background quasars have been identified. These quasars provide a quasi-inertial reference
frame with respect to which (in conjunction with existing radial velocity measurements) the absolute total space velocity vector of the
SMC can be precisely determined (see e.g. Pedreros, Costa \& M\'endez 2006). The imaging strategy was therefore
designed in a way that would satisfy the needs of the
stellar populations program as well as the
astrometry program. Given that for the latter it was mandatory to obtain {\it R}-band images, this bandpass together with  {\it B}
 imaging
was chosen to construct the CMDs.

The observations were made with a 24 $\mu$m/pixel Tektronic 2048x2048 CCD detector attached to the Cassegrain focus of
the 100-inch Ir{\'e}n{\'e}e du Pont telescope (C100) at Las Campanas Observatory, in Chile. This combination
gives a field size of 8.85$\arcmin$$\times$8.85$\arcmin$ and a scale of 0.259$\arcsec$/pixel.
 Inverse gain was 3e$^{-}$/ADU with a read noise of 7e$^{-}$.
Basic reduction of the CCD frames was done using standard IRAF\footnote{version 2.11.3, NOAO, University of Arizona} tasks.
For this purpose,
bias exposures, sky flats, and dome flats were taken every night.

\begin{figure}
\plotone{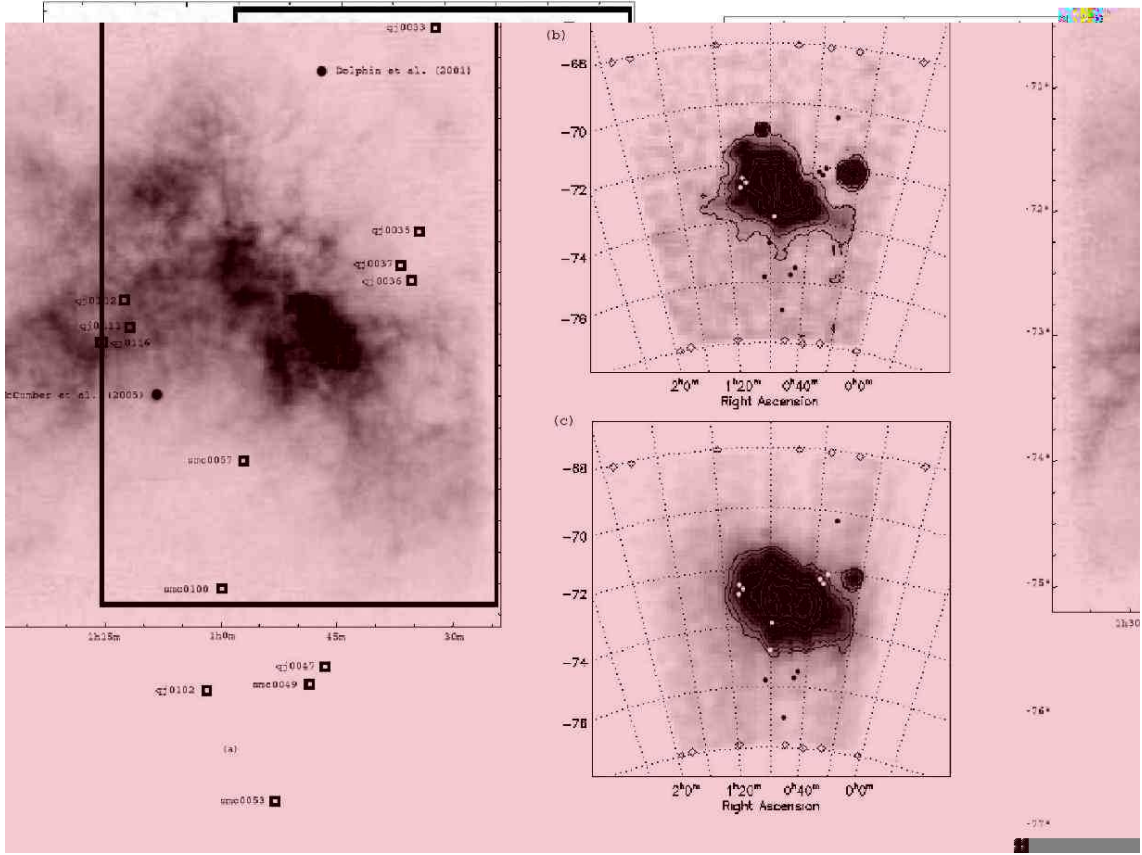}
\caption{Panel (a) shows the
spatial distribution of our SMC fields (black squares) overlapped on an HI column-density image of the SMC from 
Stanimirovi\'c et al. (1999). The big rectangle denotes the area covered by the MCPS (Zaritsky et al. 1997).
The position of the fields studied by Dolphin et al.
(2001) and McCumber et al. (2005) is also superimposed.
 The actual figure from Stanimirovi\'c et al. (1999) is delimitated
by the labelled box and includes the range in declination from $\sim$-75{\degr} to  $\sim$-70{\degr}.
 The grey-scale intensity range is 0 to 1.03$\times$10$^{22}$ atom/cm$^{2}$
 with a linear transfer
function. The maximum HI column-density is 1.43$\times$10$^{22}$ atom/cm$^{2}$.
 Panels (b) and (c), are taken from Cioni et al. (2000), and show the  distribution of young and RGB stars in the  SMC respectively.
 The position of our fields has also been overlapped.
  The fields  superimposed are the same in the three figures but the name of each  field
 in panels (b) and (c) have been omitted for clarity. 
 \label{SMCHI}}
\end{figure}

Throughout our four year campaign (2001-2004), {\it B} and {\it R} band images of thirteen fields in the SMC were obtained.
 The
coordinates of these fields and the data obtained for each of them are detailed in Table~\ref{fields}, where the first column denotes the
field, the second and third columns the right ascension and the declination respectively, fourth and fifth
columns the galactocentric distance
 (r) and the position angle (p), and the sixth and seventh columns the integration times in {\it B} and {\it R}.
  Seeing was typically
between 0.$\arcsec$7 and 1.$\arcsec$2 in all epochs.
Figure~\ref{SMCHI}a shows a HI column-density image of the SMC taken from Stanimirovi\'c et al. (1999) with
 our SMC fields (black squares) and the fields from Dolphin et al. (2001) and McCumber et al. (2005) (black circles)
 overlapped. The black rectangle denotes the area covered by the MCPS (Zaristky et al. 1997).
 The actual figure from Stanimirovi\'c et al. (1999) 
 includes the range in declination from $\sim$-75{\degr} to  $\sim$-70{\degr}. 
 Those squares denoted with ``qj'' represent the fields centered in quasars observed for astrometric purposes. We followed the
 naming convention by Tinney et al. (1997), in which the numbers following ``qj'' refer to the right ascension of the quasar (the
 declination, includded in the full
 Tinney naming convention, was ommitted).  
 The fields denoted with ``smc'' (and followed by the right ascension, for consistency with the naming convention) 
 were specifically selected for this project and were chosen to span a wide range of galactocentric distances, from
 $\sim$1${\degr}$ to $\sim$4${\degr}$.
 The combination of ``qj'' and ``smc''
   fields results in a sampling of the SMC that covers both a good range of galactocentric distances and position angles.
  It is important to note that given the elongation of the SMC, with a position angle of
 $\sim$45{\degr} (Cioni et al. 2000), a certain galactocentric distance in the Northwest or 
 Southeast directions is different than in the Southwest or Northeast 
 directions in which the stellar densities are larger. 
  Figure~\ref{SMCHI}b show the distribution of young MS stars
  and Figure~\ref{SMCHI}c the layout of the RGB (intermediate-age and old population) stars in
 the SMC from Cioni et al. (2000) with our SMC fields superimposed. The name
  of each field was omitted for clarity.
 
 \begin{deluxetable}{rrrrrrrrr}
\tabletypesize{\scriptsize}
\tablewidth{0pt}
\setlength{\tabcolsep}{0.05in}
\tablecaption{Data obtained in the SMC}
\tablehead{
\colhead{Field} & \colhead{$\alpha_{2000}$} & \colhead{$\delta_{2000}$} & r(\arcmin)\tablenotemark{a}& p({\degr}) & \colhead{{\it B}-band exposures (sec)} &
\colhead{{\it R}-band exposures (sec)} }
\startdata

smc0057& 00:57 & -73:53 & 65.7 & 164.4 &1$\times$60+1$\times$600+2$\times$800+2$\times$1000+1$\times$1200 & 2$\times$60+2$\times$600+3$\times$800+1$\times$1200\\  
qj0037 & 00:37 & -72:18 & 78.5& 294   &1$\times$60+5$\times$800&  7$\times$60+16$\times$600\\ 
qj0036 & 00:36 & -72:25 & 79.8 & 288  &2$\times$60+10$\times$600+12$\times$800& 15$\times$60+26$\times$500+22$\times$600\\
qj0111 & 01:11 & -72:49 & 80.9 & 89.5   &1$\times$60+3$\times$800&  8$\times$60+4$\times$500+17$\times$600+1$\times$700+1$\times$800\\
qj0112 & 01:12 & -72:36 & 87.4 & 81   &1$\times$60+1$\times$600+6$\times$800&  9$\times$60+7$\times$500+15$\times$600\\
qj0035 & 00:35 & -72:01 & 95.5 & 300.6  &4$\times$600+1$\times$700& 4$\times$60+1$\times$100+3$\times$500+15$\times$600+3$\times$800\\ 
qj0116 & 01:16 & -72:59 & 102.5 & 95.2  &1$\times$60+13$\times$600+2$\times$800&  7$\times$60+14$\times$500+20$\times$600\\ 
smc0100& 01:00 & -74:57 & 130.4 & 167.5 &1$\times$60+7$\times$800+3$\times$900 &1$\times$60+7$\times$600+3$\times$700\\ 
qj0047 & 00:47 & -75:30 & 161.7 & 187.7 &1$\times$800+3$\times$1000&  6$\times$60+16$\times$600+2$\times$800\\ 
qj0033 & 00:33 & -70:28 & 172.9 & 325  &1$\times$60+5$\times$600+1$\times$700& 6$\times$60+3$\times$500+20$\times$600\\ 
smc0049& 00:49 & -75:44 & 174.8 & 184.6 &1$\times$60+1$\times$600+4$\times$800 & 1$\times$60+5$\times$600\\ 
qj0102 & 01:02 & -75:46 & 179.5 & 169.4 &3$\times$60+6$\times$800&  8$\times$60+3$\times$500+15$\times$600+1$\times$700+2$\times$800\\ 
smc0053& 00:53 & -76:46 & 236.3 & 179.4 &1$\times$60+8$\times$800+2$\times$900& 1$\times$60+8$\times$600+2$\times$700\\ 

\enddata
\tablenotetext{a}{Distance from the SMC center, $\alpha_{2000}=$ 00:52.7, $\delta_{2000}=$ -72:49}
\label{fields}
\end{deluxetable}

\section{THE PHOTOMETRY} \label{photometry}

\subsection{Photometry of SMC fields}

We performed our photometry using the DAOPHOT II/ALLSTAR software
package (Stetson 1987, and updates: Stetson 1990, 1992) to find stars and to determine
an empirical PSF for each frame and then simultaneously fit it
 to all detected stars.
After testing various PSF models, we adopted a Moffat function with $\beta$=2.5 (Stetson et al. 2003)
as the best analytical model. The Moffat function is usually a much better representation of a star image than a
Gaussian function, because it has more prominent and extended wings, like real star images. Also, unlike other
functions (e.g. the Lorentz function), the Moffat function offers the option of tuning the
radial falloff of the wings to match observations.
 Between 50 and 100 stars were used
to construct the PSF of each image. Quadratic spatial variation was included because it reduced the fitting errors.

After the DAOPHOT/ALLSTAR analysis
of all the original frames was performed, an optimum, complete star list was achieved by cross matching the ALLSTAR
 results of the individual frames
with DAOMASTER (Stetson 1993).
 ALLFRAME (Stetson 1994) was then used to obtain simultaneous photometry of stars in all CCD images of each SMC field.

After  ALLFRAME, we ran MONTAGE2 for each field.
 MONTAGE2 takes the images from which all the stars
have been subtracted according to their final positions and brightnesses, and applies the known geometric transformations that relate their
coordinate systems and the known differences in their photometric zero-points, to produce a median image of each field.
We tested in each of the 13 SMC fields that the number of stars detected on the median of the subtracted images, 
and considered real stars after
running DAOPHOT, ALLSTAR, and ALLFRAME a second time, was less than 1\%, i.e. negligible for our goal.

ALLFRAME produces an image-quality index, CHI, which is a dimensionless measure of the agreement between the 
brightness profile of any given object and that of the model PSF for the frame in which it is measured. Similarly,
ALLFRAME gives the SHARP index, which is a first-order estimate of the intrinsic angular radius of a source.
Another index provided by ALLFRAME is $\sigma$, which represent the standard error
 of the star's magnitude and is representative of the internal errors of the photometry.
 Stars with at least one valid measurement in each band ({\it B} and {\it R}) were selected, and their final magnitudes were obtained
 using DAOMASTER, which combines the magnitudes measured for each star in each image to provide the ``mean weighted'' magnitude. DAOMASTER
 also provides $\sigma$, CHI, and SHARP parameters for each stars, which are a combination of the corresponding parameters for each image.
 We used the
following limits for the error and shape parameters
given by DAOMASTER: $\sigma$$_{(B-R)}$$^{2}$=$\sigma$$_{B}$$^{2}$+$\sigma$$_{R}$$^{2}$$\leqslant$0.15,
-0.6$\leqslant$$\mid$SHARP$\mid$$\leqslant$0.6,  and $\mid$CHI$\mid$$\leqslant$2.5.
 The final number of stars we kept, measured in {\it (B-R)} of each field
(and consequently, in each CMD)
 is given in Table~\ref{exposure}. A total of 215,121 stars down to {\it R}$\thickapprox$24
 were measured.
 
 \begin{deluxetable}{ccccccccccccccccc}
\tabletypesize{\scriptsize}
\tablewidth{0pt}
\setlength{\tabcolsep}{0.3in}
\tablecaption{Final number of stars measured in {\it (B-R)}}
\tablehead{
\colhead{Field} & \colhead{N\tablenotemark{\dag}}}
\startdata

smc0057 &  21246 \\ 
qj0037  &  21579 \\
qj0036  &  23756 \\ 
qj0111  &  25845 \\ 
qj0112  &  47548 \\
qj0035  &  24114 \\    
qj0116  &  19875 \\
smc0100 &   7801 \\
qj0047  &   9762 \\ 
qj0033  &   3797 \\ 
smc0049 &   3580 \\ 
qj0102  &   4030 \\
smc0053 &   2188 \\

\label{exposure}
\enddata
\tablenotetext{\dag}{Selected stars with $\sigma$$\leqslant$0.15,\\ CHI$\leqslant$2.5 and -0.6$\leqslant$SHARP$\leqslant$0.6.}

\end{deluxetable}

The so-called
aperture corrections, i.e., the corrections that place the relative
 profile-fitting magnitudes on the system of the ``total'' instrumental magnitudes for a particular frame,
 were obtained from synthetic aperture photometry by measuring several isolated, bright stars through a series of increasing
 apertures and the
  construction of growth curves (Stetson 1990).
  We used a growth sequence which consists in twelve apertures from {\it $r_{1}$} to {\it $r_{12}$}
in such a way that we considered the first radius {\it $r_{1}$} about half of the one we used for the aperture photometry of our objects
 and {\it $r_{12}$}=30$\arcsec$. The sequence is {\it $r_{k}$=$\sqrt[11]{r_{12}/r_{1}}$$\times$ $r_{k-1}$ }, {\it $k=2,...,12$}.
  The aperture corrections were derived using the program DAOGROW (Stetson 1990). In each image we selected
  the brightest and most isolated 30 to 80 stars among those used in the
   derivation of the
  PSF for that frame. The
  aperture photometry results were then provided to DAOGROW which returns the ``total'' instrumental magnitude
  and its standard error for each of the selected stars. The aperture correction for
  a particular frame was obtained as the median of the differences
  between the ``total'' magnitude and the profile-fitting ALLFRAME magnitude for all
  selected stars on that frame. Errors in the aperture corrections were calculated as the
  standard error of the mean of these differences, and were typically between $\pm$0.001 and $\pm$0.003.

 \subsection{Standard Stars Photometry}\label{calibration}

Our instrumental photometric system was defined by the use of the Harris {\it BVRI} filter set, which constitutes the default
option on the C100 for broad-band photometry on the standard Johnson-Kron-Cousins system. In photometric nights, typically six
{\it BVRI} standard star areas from the catalog of Landolt (1992) were observed several times to determine the transformation of our
instrumental magnitudes to the standard {\it BR} system. Thirty-one different standard stars with
 colors: -0.49$<${\it (B-R)}$<$5.00
 were observed, and a total of 272 observations were made; 178 in 2001 and
  94 in 2002.
Most of these areas include stars of a wide variety of colors. A few of
 them were
followed up to about 2.0 airmasses to determine atmospheric extinction optimally. During these nights we also obtained
 short exposure {\it BR} frames of all our fields of interest, which served the purpose of calibrating all frames
taken with non-photometric sky. Field qj0035 is not considered in this paper because we do not have yet a secure calibration for it.

 The Landolt standard images are uncrowded fields, thus no profile-fitting photometry was necessary for them.
   Instead,  DAOGROW
 was used in an identical way as for the program stars to directly derive the total instrumental magnitudes
 for the standard stars from their aperture photometry.

To put our observations into the standard system, we used the following transformation equations:

\begin{eqnarray} \label{transf}
\nonumber B=b+\alpha_{b}+\beta_{b}(B-R)+\gamma_{b}X_{b} \\
 \\
\nonumber R=r+\alpha_{r}+\beta_{r}(B-R)+\gamma_{r}X_{r}
\end{eqnarray}

 where $(b,r)$ and $(B,R)$ are the instrumental
and standard magnitudes respectively, and (X$_{b}$, X$_{r}$) are the airmasses.
No time dependence terms were added since a preliminary fit showed no trends in the residuals of both {\it B} and {\it R} with time.

The color terms
($\beta$$_{b}$, $\beta$$_{r}$) and extinction  coefficients ($\gamma$$_{b}$, $\gamma$$_{r}$),
as well as the zero-points ($\alpha$$_{b}$, $\alpha$$_{r}$) are
 unknown, presumably constant, transformation coefficients and are to be calculated.  Both the color-dependent term and the zero-point
  in the
transformation, are expected to be reasonably constant properties of the telescope/filter/detector
combination and, in fact, the night-to-night differences for a given year
 in their computed values were found to be consistent with the uncertainty of the
individual determinations.
The above equations  were applied to the Landolt standard star magnitudes, and solved for
$\alpha$$_{b}$, $\alpha$$_{r}$, $\beta$$_{b}$, $\beta$$_{r}$, $\gamma$$_{b}$ and $\gamma$$_{r}$, for each night, using a custom program.

\begin{deluxetable}{ccrrrrrrrrcrl}
\tabletypesize{\scriptsize}
\tablewidth{0pt}
\setlength{\tabcolsep}{0.3in}
\tablecaption{Transformation equation coefficients}
\tablehead{
\colhead{Campaign} & \colhead{$\alpha$$_{b}$} & \colhead{$\alpha$$_{r}$} & \colhead{$\beta$$_{b}$} & \colhead{$\beta$$_{r}$} 
}
\startdata

2001  &  -0.709$\pm$0.002 & -0.423$\pm$0.001 & +0.040$\pm$0.002 &  +0.001$\pm$0.001 \\ 
2002  &  -0.892$\pm$0.002 & -0.549$\pm$0.002 & +0.045$\pm$0.002 &  +0.002$\pm$0.002 \\

\label{coefficients}
\enddata
			   
\end{deluxetable} 

\begin{figure}
\plotone{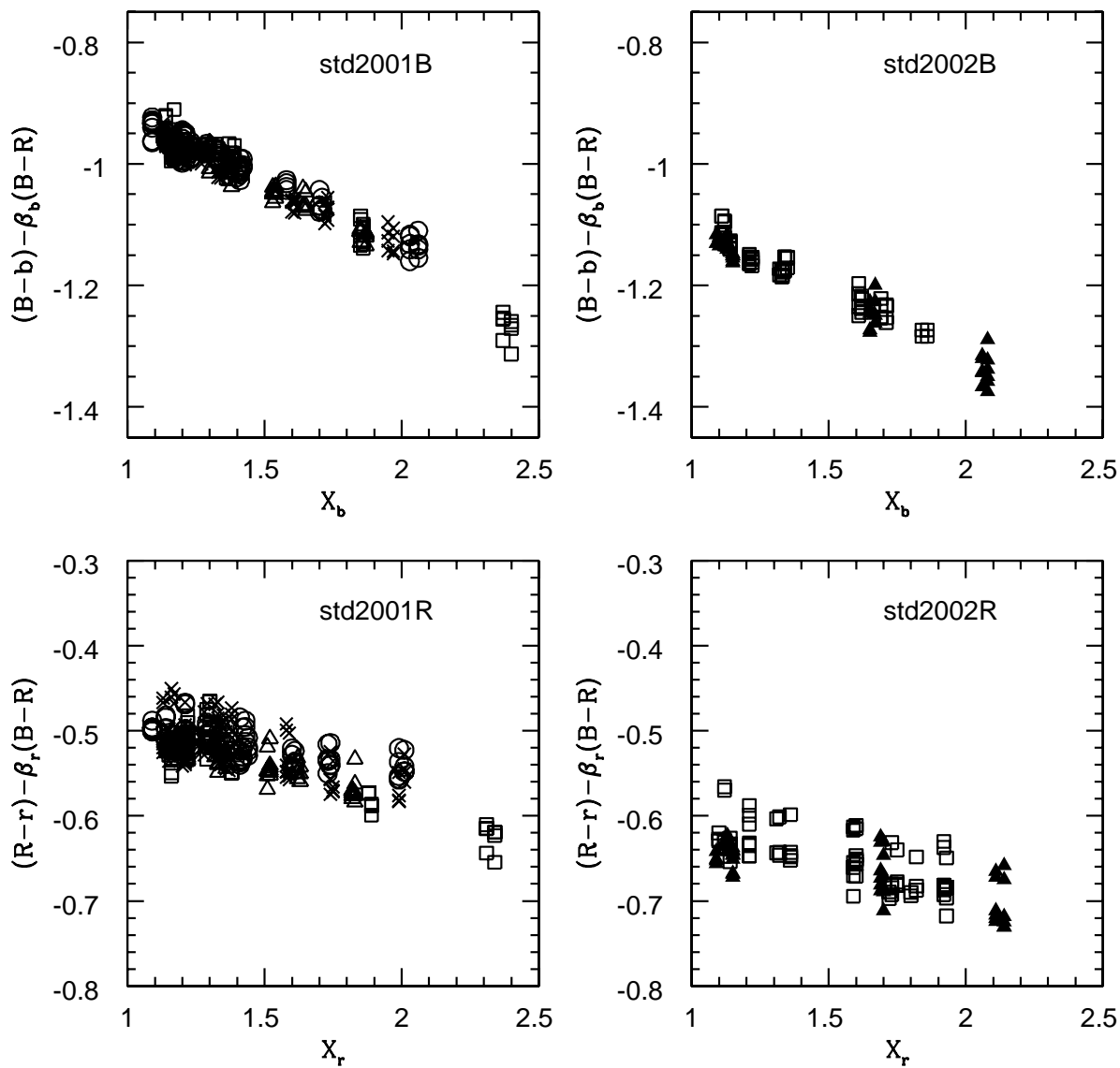}
\caption{The upper panels show the [B-b-$\beta$$_{b}$(B-R)] vs. X$_{b}$ (for standards in {\it B}-band) 
and the lower panels show the 
[R-r-$\beta$$_{r}$(B-R)] vs. X$_{r}$ (standards in {\it R}-band) for 2001 and 2002. 
The slope correspond to the extinction coefficient value. Open triangles correspond to 10/16/2001, open squares to 10/17/2001, 
 diagonal crosses to 10/18/2001 and open circles to 10/19/2001. For 2002, open squares correspond to 10/10 and filled 
  triangles to 10/13.\label{four}}
\end{figure}

Then, we followed an iterative procedure to refine our photometric transformation. First,
a new set of unique ($\alpha$$_{b}$, $\alpha$$_{r}$) and ($\beta$$_{b}$, $\beta$$_{r}$) values for each campaign was obtained by
 imposing the
extinction coefficients ($\gamma$$_{b}$, $\gamma$$_{r}$) corresponding to each night. Then, we applied the resultant zero-points and
color terms
 to each night of each year and new extinction coefficients were derived for each night. In this way, 
we have a set of ($\alpha$$_{b}$, $\alpha$$_{r}$) and ($\beta$$_{b}$, $\beta$$_{r}$) values for each year
 and ($\gamma$$_{b}$, $\gamma$$_{r}$) values
for each night of each year.
In Table~\ref{coefficients} we present the zero-points and the color coefficients. The first column denotes the year, the
 second and third columns indicate the ($\alpha$$_{b}$, $\alpha$$_{r}$) coefficients,
  and fourth and fifth columns the ($\beta$$_{b}$, $\beta$$_{r}$) coefficients.
  The final values for
the  extinction coefficients are shown in the second and third columns of Table~\ref{atmospheric_extinction}, where the first column
shows the night to which the coefficients correspond.
Figure~\ref{four} shows the standard distributions for each filter and each year; the slope represents the extinction
coefficient.
Fitting errors in the zero-points correspond to $\sigma$/$\sqrt{(N-2)}$, where N is the number of measurements we have.
 They turned out to be between $\pm$0.001 and $\pm$0.002. 
 
The total zero-point errors of the photometry, including the error in the extinction, in the aperture corrections, and
 the uncertainties in the calibrations, are $\sim$0.02 mag in both {\it B} and {\it R}. These values are consistent with the
 systematic errors derived comparing photometry of different epochs.
 Using those SMC fields for which we have standard star observations from more than one night, we
 tested the photometric differences resulting from the use of different standard sets
 corresponding to different nights. We found that these  
differences are between $\pm$0.001 and $\pm$0.03 in {\it B} and between $\pm$0.001 and $\pm$0.04 in {\it R}
 (standard errors of the mean).

\begin{deluxetable}{ccc}
\tabletypesize{\scriptsize}
\tablewidth{0pt}
\setlength{\tabcolsep}{0.3in}
\tablecaption{Atmospheric extinction coefficients}
\tablehead{
\colhead{Night} & \colhead{$\gamma$$_{b}$} & \colhead{$\gamma$$_{r}$} 
 }  

\startdata

16/10/2001  & -0.218$\pm$0.007 &     -0.078$\pm$0.003 \\  
17/10/2001  & -0.219$\pm$0.007 &     -0.077$\pm$0.007 \\  
18/10/2001  & -0.219$\pm$0.006 &     -0.070$\pm$0.009 \\  
19/10/2001  & -0.211$\pm$0.005 &     -0.065$\pm$0.005 \\ 
10/10/2002  & -0.207$\pm$0.008 &     -0.066$\pm$0.009 \\ 
13/10/2002  & -0.215$\pm$0.008 &     -0.076$\pm$0.008 \\ 

\enddata
\label{atmospheric_extinction}

\end{deluxetable}

\section{THE SMC STELLAR CONTENT} \label{stellar}

We present unprecedented deep SMC CMDs obtained using a medium aperture telescope (100-inch). The quality of the CMDs
is comparable to that of the CMDs obtained for the MCs using the HST (e.g. McCumber et al. 2005).
Given the depth  of our diagrams, which reach the oldest turnoffs, accurate information about age can be derived from the
 MS turnoff luminosities. As discussed by Gallart, Zoccali, \& Aparicio (2005), reaching the oldest turnoffs will allow us to break the
 age-metallicity degeneracy affecting most methods to obtain the SFH.
 
\begin{figure}
\plotone{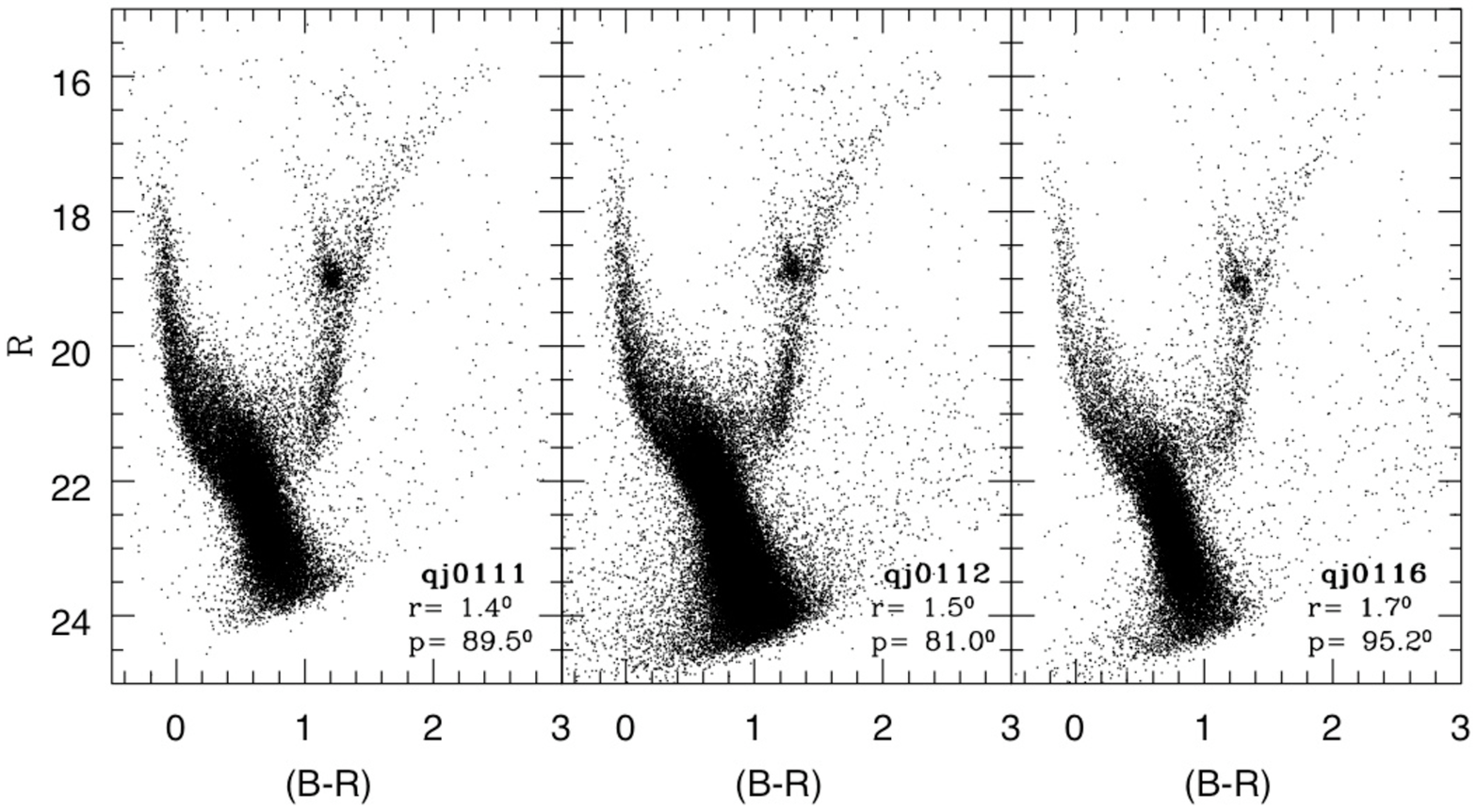}
\caption{CMDs corresponding to the fields on the Eastern side of the SMC, in order of
increasing distance from the SMC center (r) and with different position angles (p).
\label{three_east}}
\end{figure} 
\begin{figure}
\plotone{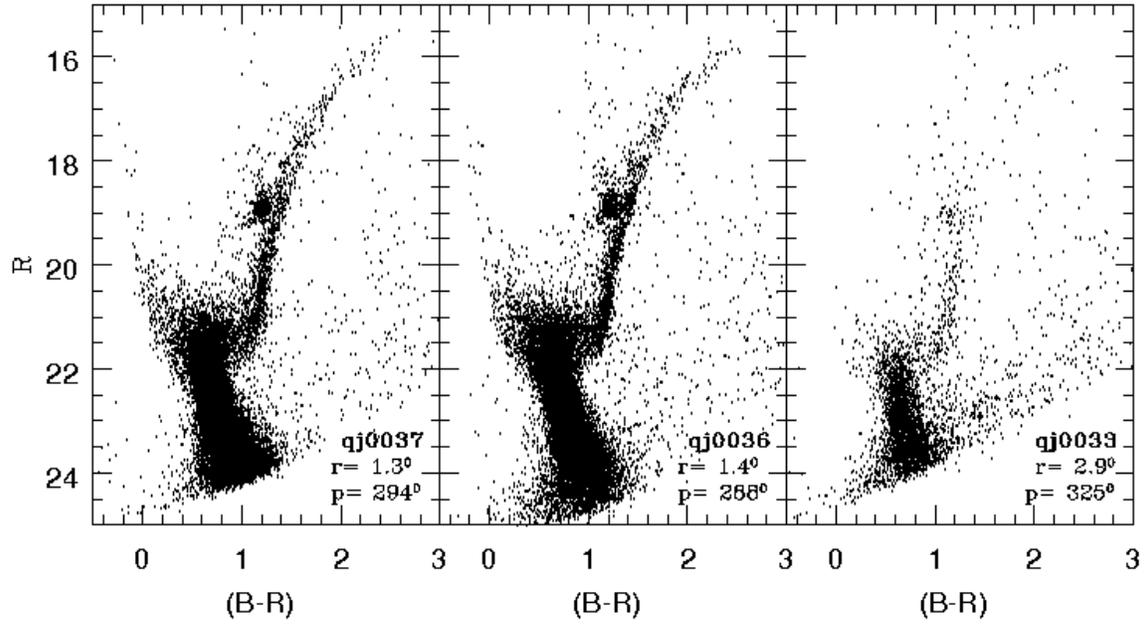}
\caption{CMDs corresponding to the Western side of the SMC, in order of 
increasing distance from the SMC center (r) and for different position angles (p).
\label{three_west}}
\end{figure}
\begin{figure}
\plotone{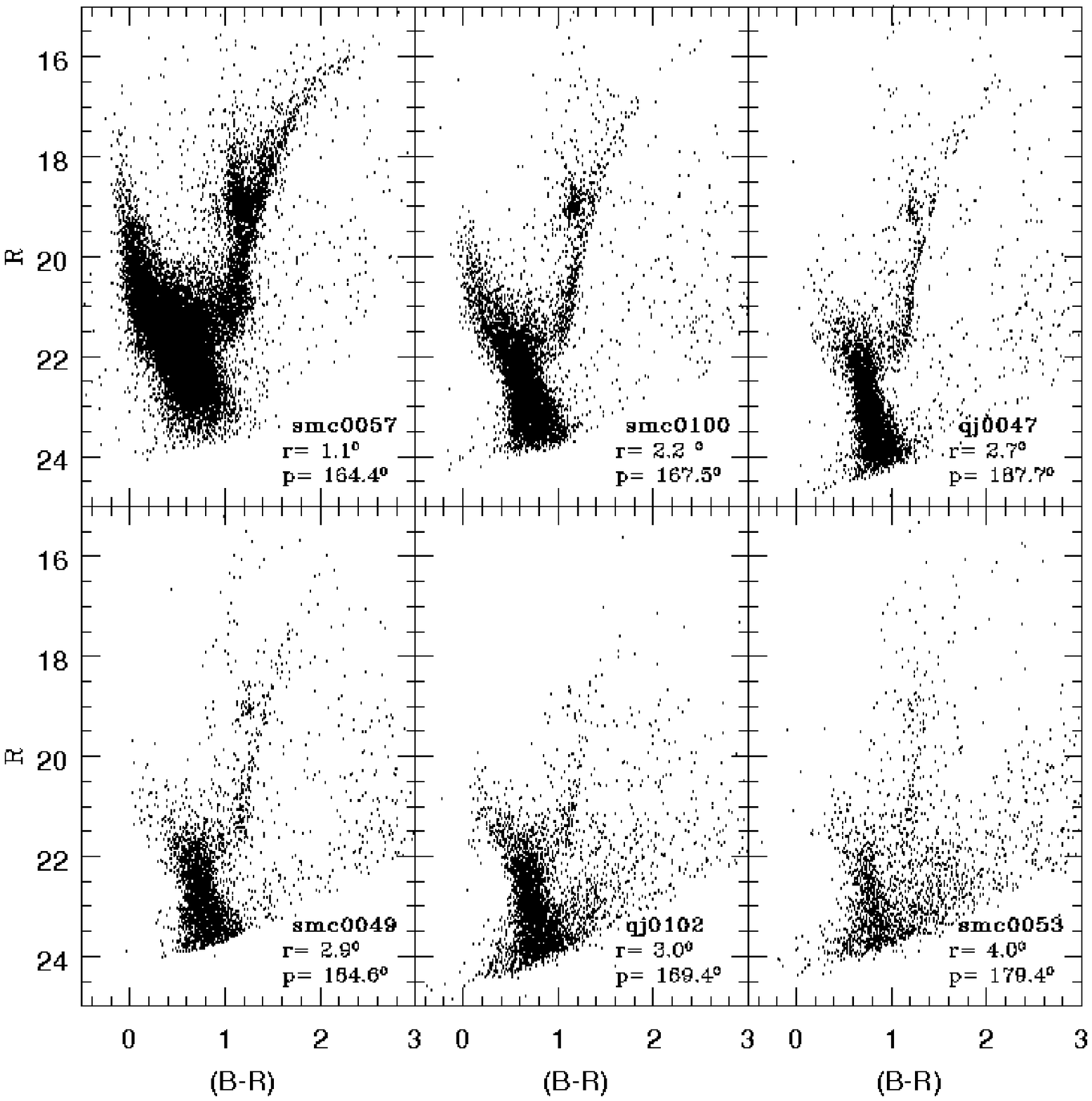}
\caption{CMDs corresponding to the Southern side of the SMC, in order of
increasing distance from the SMC center (r) with different position angles (p).
\label{six_south}}
\end{figure}
 
Figures~\ref{three_east} to ~\ref{six_south} show the [{\it(B-R),R}] CMDs for the twelve
SMC fields in order of increasing galactocentric distance.
 The fields range from galactocentric radius $\sim$1$\degr$ to $\sim$4$\degr$, and sample different
 regions of the SMC, as shown in Figure~\ref{SMCHI}, where the fields are depicted in the context of the HI distribution
 (Figure~\ref{SMCHI}a), and of the young (Figure~\ref{SMCHI}b) and intermediate-age and old (RGB stars)
   populations (Figure~\ref{SMCHI}c).
All the CMDs reach a couple of magnitudes below the old MS
turnoff (even the most populated), with great photometric accuracy.
 
  Figure~\ref{three_east} shows the CMD of fields qj0111,
qj0112, and qj0116,
 which have a galactocentric radius ranging from
  1.4$\degr$ to 1.7$\degr$. These fields correspond to the Eastern side of the galaxy (facing the LMC) as seen in Figure~\ref{SMCHI}a,
 from which we can notice that they are located in an area of high HI concentration, inside the Supergiant Shell 304A with HI mass of 
  5.7$\times$10$^{7}$ M$_{\odot}$ (the total HI mass in the area observed by Stanimirovi\'c et al. is $\sim$4$\times$10$^{8}$
   M$_{\odot}$).
  These fields are also in the region of the bridge which connects the so-called ``stellar bar''\footnote{The SMC
   is not a type of barred Irr galaxy, for which the LMC is the prototype,
   but sometimes it is useful to call bar to the brightest portion of its chaotic major axis.} with the ``Wing''. From
  Figure~\ref{SMCHI}b and Figure~\ref{SMCHI}c, it can be noticed that these fields are located
  in very densely populated isopleths of young, intermediate-age and old population. 
  Indeed, the
    CMDs of these fields show a conspicuous MS, well populated from the oldest turnoff at R$\sim$22 to  R$\sim$16,
 which implies the presence of a large number of very young stars in all the CMDs.
  The three Western fields are also shown in 
  Figure~\ref{SMCHI}a and the corresponding CMDs are presented in order
   of increasing galactocentric distance from 1.3$\degr$ to 2.9$\degr$ 
  in Figure~\ref{three_west}. These fields are
  situated in the opposite side of the LMC, in a region with low concentration of HI. From Figure~\ref{SMCHI}b it 
  is evident that they are located in an area of low young stellar density,
   but still high density of intermediate and old population similar to that of the Eastern fields. 
The CMDs of these Western fields show a less significant MS, even at similar distances from the center as the Eastern fields. This 
fact is 
in agreement with the low density of the HI column in this part of the galaxy (see Figure~\ref{SMCHI}a).
We would like to address if the differences seen in the young population can be extended to the intermediate and old population.

  Figure~\ref{six_south} shows the CMDs of the six Southern fields, whose galactocentric distances range from
 1.1$\degr$ to 4$\degr$. As seen in Figure~\ref{SMCHI}a these fields are located in regions in which the HI column-density is very low.
 Field smc0057,  the  closest to the center, presents a conspicuous young MS.
  In Figure~\ref{SMCHI}b is shown that the stellar density of young stars in this field 
 is higher than in the rest of the Southern fields and similar to the Eastern field qj0111. Field smc0100 still shows many
 bright young MS stars but further on 2.7$\degr$ 
 the CMDs have less populated young MS, and  present a similar distribution of stars (with less statistics while going 
 further South). The four furthest fields are located in an area in which the stellar density is similar, as shown in
 Figures~\ref{SMCHI}b and~\ref{SMCHI}c.
 
Depending on their mass and metallicity, the core He-burning stars produce the horizontal branch (HB) and the
 RC.
 The presence of a prominent blue HB
points out the existence of an old, metal poor population; the lack of such HB and the presence of a RC
may indicate that core He burners are not so old,
 or more metal rich, or both (see Gallart et al. 2005 for details).
  A shared feature in all our SMC CMDs is the absence of a well populated, blue extended HB, pointing out that the amount
   of
  field stellar population as old and metal poor as that of the Milky Way halo globular clusters and dwarf spheroidal
  galaxies is small in the SMC.
  The RC is well populated and 
extended in luminosity, with a width of up to
$\triangle$R$\sim$1 mag, denoting the presence of a large intermediate-age population.
The fields with a prominent bright MS, as the Eastern fields and smc0057, for example, show the  \emph{vertical extension of the RC}
feature corresponding to non-degenerate core He-burners (Gallart 1998).
The
older stars in the core He-burning phase lie in the lower part of the observed RC, while younger stars are brighter.

In the following sections, we discuss the main features of the
SMC CMDs using isochrones (sec \S~\ref{isochrones}) and Color Functions (sec \S~\ref{CF}). This will
allow us to describe the SMC stellar
populations, as a starting point to a study (No\"el et al., in preparation) in which
we will address the SMC SFH by comparing in detail the distribution of stars in
the observed CMDs with a set of model CMDs.

\subsection{Theoretical isochrones} \label {isochrones}

\begin{figure}
\plotone{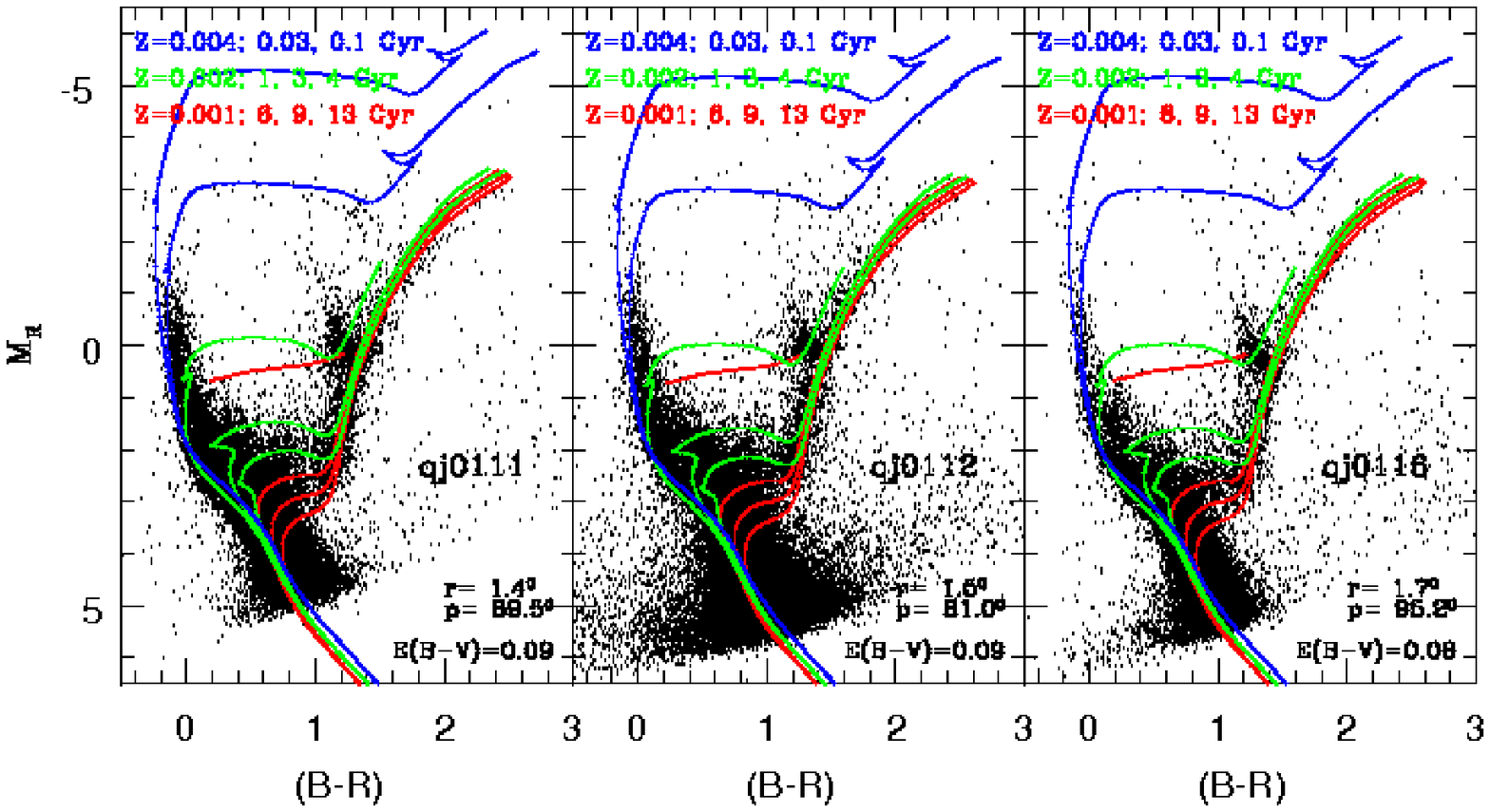}
\caption{Fields located to the East side of the SMC, 
with Teramo isochrones (Pietrinferni et al. 2004) 
superimposed. The CMDs were transformed to absolute magnitudes using 
(m-M)$_{0}$=18.9, and de-reddened with the reddening labelled, obtained using the criteria explained in the 
text.\label{three_east_iso}}
\end{figure}

\begin{figure}
\plotone{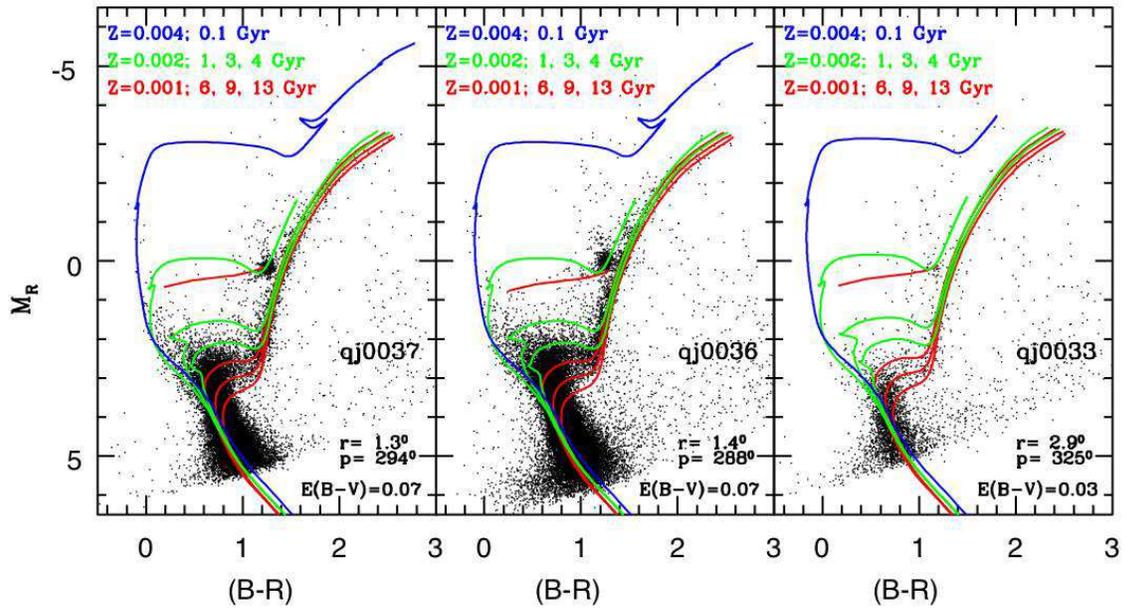}
\caption{As in Figure~\ref{three_east_iso}, but showing the CMDs of fields located to the West  side of the SMC.\label{three_west_iso}}
\end{figure}

\begin{figure}
\plotone{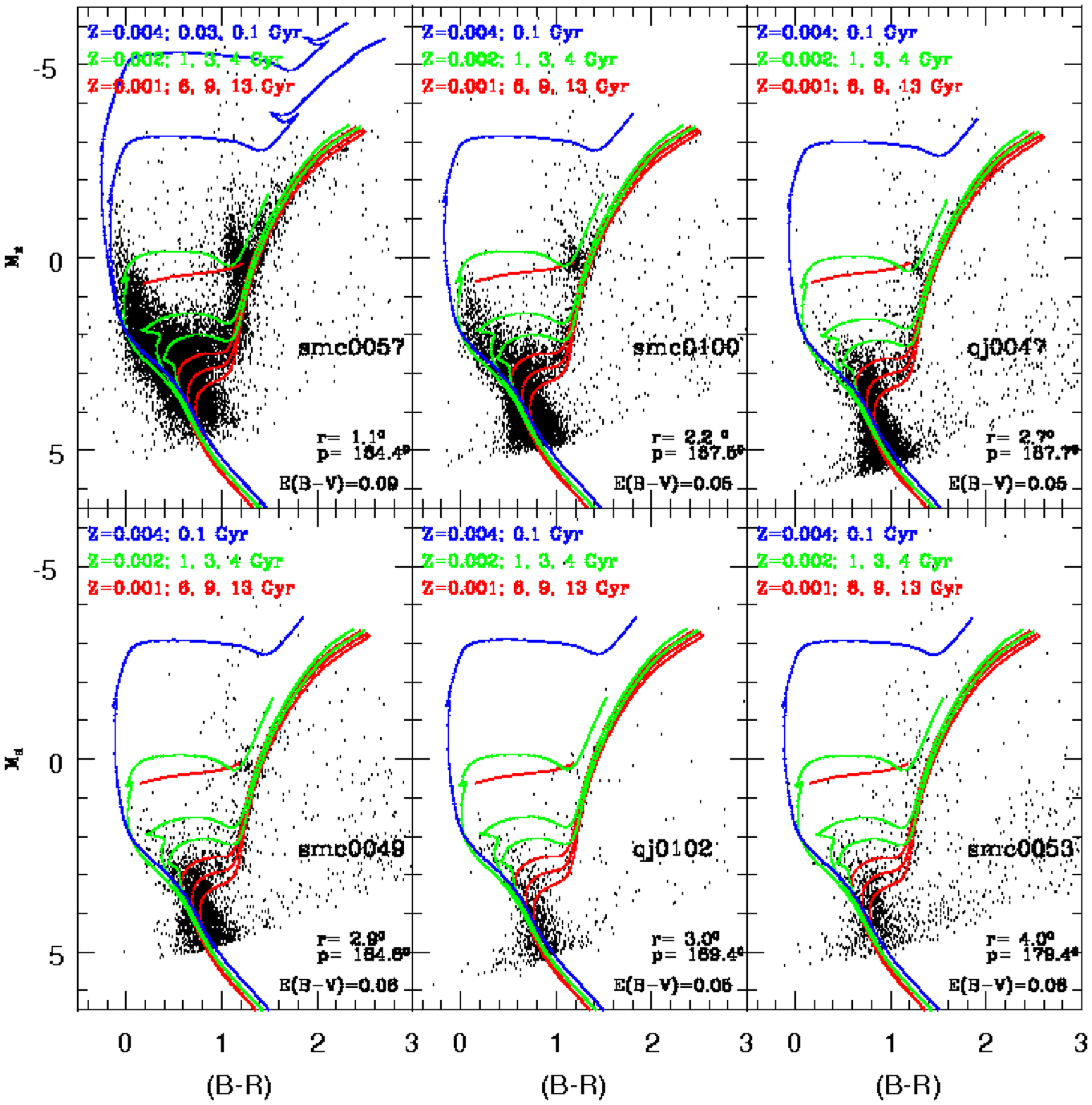}
\caption{As in Figure~\ref{three_east_iso}, but showing the CMDs of fields located to the South of the SMC.\label{six_south_iso}}
\end{figure}

The interpretation of CMDs of composite stellar population strongly relies on the stellar evolution models adopted (see
Gallart et al. 2005). For the present work, we used the Teramo stellar evolution models (Pietrinferni et al. 2004) as the reference.
 In Figures~\ref{three_east_iso}, ~\ref{three_west_iso}, and  ~\ref{six_south_iso} 
we have superimposed Teramo isochrones to our CMDs for three different metallicities suitable
for the SMC stellar populations: Z=0.001 ([Fe/H]=-1.27), Z=0.002 ([Fe/H]=-0.96), and Z=0.004 ([Fe/H]=-0.66).
The final selection of the metallicities was done taking into account the RGB and MS star colors and its best match by Teramo isochrones 
of given ages and metallicities.

    
The observed CMDs have been transformed to absolute magnitudes
using a distance modulus of $(m-M)_0=18.9$ (see van den Bergh 1999) and de-reddened as follows.
 For the two outermost fields, the reddening values given by the IRAS/COBE extinction maps (Schlegel, Finkbeiner \&
 Davis 1998) were used. For the innermost regions, Schlegel et al. (1998) estimate a typical reddening
 of E(B-V)=0.037, from the median dust
emission in surrounding annuli, since reddenings {\it through} the SMC cannot be calculated because its temperature
structure is not sufficiently well resolved by DIRBE. Their quoted value, therefore, is not accurate enough and we have estimated a mean
value of the reddening for each of the fields smc0057, qj0037, qj0036, and qj0111 by requiring a good fit
of the CMD by the same isochrones that were overlapped to the other fields,
 for which the IRAS/COBE reddening values are
reliable. This assumes the same age-metallicity relation for all fields, which is in agreement with the results found by Carrera (2006)
and Carrera et al. (2006, in preparation) using CaII triplet spectroscopy.

In all cases, the relations 
$A_B=1.316$E(B-V), and $A_R=0.758$ E(B-V) (Schlegel et al. 1998) have been assumed to calculate the
extinction in the {\it B} and {\it R} bands.

 Canonical isochrones of 6, 9, and 13 Gyr with Z=0.001, isochrones with overshooting of 1, 3, and 4 Gyr with Z=0.002,
 and of 0.1 Gyr with Z=0.004 were overlapped. In the Eastern fields and in field smc0057 a 0.03 Gyr isochrone with
 overshooting and Z=0.004 was also superimposed.
 
 Figure~\ref{three_east_iso} shows the CMDs of the Eastern fields with the superposition of Teramo isochrones.
 In these CMDs the conspicuous MS is well populated from the oldest turnoff
at M$_{R}$=3.5 (13 Gyr isochrone) up to the 0.03 Gyr isochrone. All the CMDs show a large number of young stars well represented by
the 0.03 Gyr and 0.1 Gyr isochrones. The areas around the
1 to 6 Gyr isochrones are densely populated, indicating a strong presence of
intermediate-age stars. No obvious differences between the MS of these fields
can be inferred from the comparison with isochrones alone.

  Figure~\ref{three_west_iso} shows the CMDs of the Western fields with the isochrones overlapped.
 In these fields,
the intermediate-age population ($\sim$3-4 to 9 Gyr) seems dominant.
 Fields qj0036 and qj0037 have a small fraction of stars born up to 0.1 Gyr ago,
while qj0033, the most remote of the three, seems to have no stars
younger than 1 Gyr (though this could be an effect of small number statistics).

 Figure~\ref{six_south_iso} shows the CMDs of the six Southern fields with the superposition of isochrones. 
Field smc0057, the closest to the center, presents an important MS 
 where the 0.1 Gyr isochrone is still quite populated. 
In field smc0100 there is still a substantial population between 1 and 3 Gyr old, but 
 fields  qj0047 to smc0053 show few stars 
 younger than 3-4 Gyr.
 The population of these relatively far-off fields is not purely old but contains an important amount of 
intermediate-age stars that inhabit the zones of the MS and subgiant branch 
around the 3, 4, 6, and 9 Gyr isochrones, in addition to the several 
old stars around the 13 Gyr isochrones.

\subsection{Color Functions} \label {CF}

\begin{figure}
\plotone{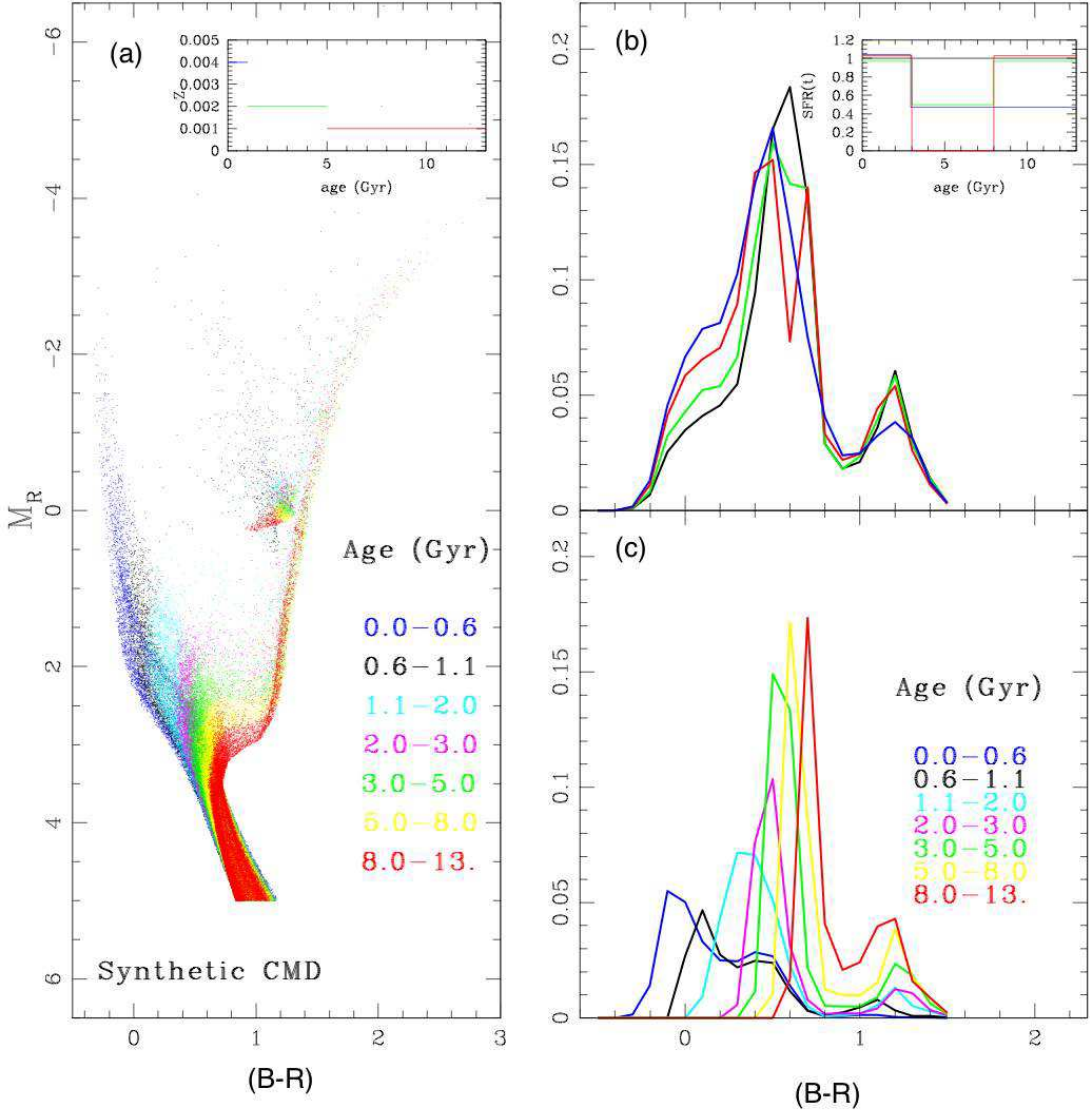}
\caption{\small{The panel (a) shows a synthetic CMD showing the position of stars in different ages intervals.
 The Teramo stellar evolution models have been used in the computation using the tool iac-STAR (http://iac-star.iac.es/iac-star/).
 Constant SFR(t) from 13 Gyr ago to
  the present time has been assumed. In the inset on top of panel (a) the metallicity law used is shown:
  Z=0.004 ([Fe/H]=-0.66) between 0 and 1 Gyr ago, 
  Z=0.002([Fe/H]=-0.96) between 1 and 5 Gyr ago, and 
  Z=0.001 ([Fe/H]=-1.27) between 5 and 13 Gyr ago. In panel (b) different CF distributions,
  corresponding to synthetic populations with different SFR(t) and  
  integrated in the range -1$<$M$_{R}$$<$3.5 are depicted (see inset).
  Black CF correspond to the CMD in panel (a). The CF for a population with an enhancement in the SFR(t) 3 Gyr ago is depicted in blue.
   The green CF represents a population in which the SFR(t) decreased by 50\% between 8 and 3 Gyr ago. The CF in red represents a
   population 
   with a gap in the SFR(t) between 8 and 3 Gyr ago.
   Panel (c) shows the CF corresponding to the CMD in panel (a) with limited age ranges, as described in the labels.\label{docfsynt}}}
\end{figure}

Gallart et al. (2005) have discussed the potential of MS Color Functions (CFs) to provide information on the stellar populations present in a galaxy.
 They concluded that, in fact, the CF may be a better tracer of the SFH than the LF. The reason is that, in CMDs like
 [{\it (B-R), R}] or [{\it (V-I), I}], the MS of isochrones of different age follow roughly vertical paths up to the MS turnoff
  where they turn
to the red, almost perpendicularly to the MS, to form the subgiant branch or, in the case of more massive stars, the blue loops. This
effect is clear from the CMD in panel (a) of Figure~\ref{docfsynt} (we will return to this figure later).
The fact that the turnoff of a stellar population gets redder as the populations gets older implies a general dependency on age of
the CF of the MS and subgiant branch stars brighter than the oldest MS turnoff. Here we will discuss CFs that will include the whole CMD
within a region comprised between: -1$\leqslant$M$_{R}$$\leqslant$3.5 and
-0.5$\leqslant$(B-R)$\leqslant$1.5. The star counts have been normalized to the total stars present in the integrated range.
 Most of the information belongs in any case to the MS and subgiant branch. The fraction of CF that belongs
to the RGB and RC serves as a check of the metallicity assumed for the synthetic CMD, and contributes to the overall normalization of the
CF.

Comparison of the stellar content
of a galaxy with isochrones and CFs is complementary in terms of the information on the stellar population they
provide.
The isochrones give us the range of ages present in a given CMD;
the comparison between theoretical and observed CFs is a good tool to assess
 the relative amounts of stars of different ages present in a stellar population.
This is a particular case of a comparison of the number of stars in a set of boxes in the observed and synthetic CMDs
respectively (e.g. Gallart et al. 1999).
 However, the CF is particularly useful for a first qualitative assessment of the SFH because it is a one-dimensional
representation of the content of the CMD in which there is a relatively direct correlation between range of ages and color.
All the models used here have been computed using the synthetic CMD algorithm iac-STAR (Aparicio \& Gallart 2004).

 Panel (a) in Figure~\ref{docfsynt} shows a synthetic CMD calculated assuming a constant SFR(t) from 13 Gyr ago to date and
    metallicities: Z=0.004 in the range 0 Gyr$\leqslant$t$\leqslant$1 Gyr, Z=0.002
     in the range 1 Gyr$\leqslant$t$\leqslant$5 Gyr, and Z=0.001 in the range 5 Gyr$\leqslant$t$\leqslant$13 Gyr (see inset).
    Stars corresponding to each age interval are depicted in different colors.
   A 30\% of binary stars have been assumed; the 
parallel feature to the red of the MS between 3.5$\leqslant$M$_{R}$$\leqslant$5 corresponds to binary stars.
 Some of these binaries between 8-13 Gyr old are noticeably brighter than single stars at the same age and mass
 (red points in the synthetic CMD between 2$\leqslant$M$_{R}$$\leqslant$3).

 Panel (b) in Figure~\ref{docfsynt} shows the composite CFs of populations with different SFR(t) (shown in the inset).
   Three main
  features characterize the CF: a local elevation in the blue part between -0.3$\leqslant$(B-R)$\leqslant$0.2, a central peak
  between 0.2$\leqslant$(B-R)$\leqslant$0.9, and a second local maximum in the red part.
 Panel (c) of Figure~\ref{docfsynt} shows which populations contribute to each feature in the CF for the synthetic CMD in panel (a).
 By comparing these, we find the following: 
the elevation in -0.3$\lesssim$(B-R)$\lesssim$0.2 in the composite CF
 mostly corresponds to stars in the age range 0$\leqslant$t$\lesssim$1.1 Gyr; in the central
 peak (0.3$\lesssim$(B-R)$\lesssim$0.9) there are populations of all ages, predominating
  the intermediate-age and old stars ($\sim$2-13 Gyr).
  The second local maximum corresponds to the RC and the red part of the subgiant branch
   and is composed by stars with age $\gtrsim$1 Gyr as seen in panel (c) in  Figure~\ref{docfsynt}.
   Going back to panel (b), the CF in black corresponds to a galaxy with a constant SFR(t) 
   (i.e., the CF of the CMD in the panel (a)). The blue CF is from a synthetic CMD computed assuming a SFR(t) that
    increased from 3 Gyr ago to the present. 
    This CF presents a higher blue elevation due to the bigger number of young blue stars; also
   the central peak is somewhat blueshifted.
     The green CF represents a stellar population in which the SFR(t)
   dropped by 
   50\% between 3 and 8 Gyr ago. This is reflected in the shape of the 
   absolute maximum, where a depression appears as a result of the
   assumed SFH.  This kind of feature would be difficult to detect in a real population.
 Finally, the red CF corresponds to a SFR(t) with a totally quiescent period between 3 and 8 Gyr ago.
The deep hollow shown in the absolute maximum of this CF is due to the gap at intermediate ages in the
 SFH. The last two CFs show a higher blue elevation as compared to a CF of a CMD with constant SFR(t) due to the higher
 fraction of young stars over the whole population.

Prior to present a discussion based on CFs, we will briefly talk about the crowding effects and, in general, the observational errors
 that are present in our photometry.
A careful study of the extent and nature of these crowding effects is key to study the
stellar populations present in the SMC. We will understand and
quantify the crowding effects using the artificial-star tests procedure.
 These artificial-star tests have been performed in a
way similar to that described in Gallart et al. (1999) and
will be presented in detail in a forthcoming paper (No\"el et al. in preparation). In short, a number of
artificial stars of known magnitudes and colors are injected in the original frames,
 using the ADDSTAR algorithm of  DAOPHOT II (Stetson 1993). The photometry is re-done following the same procedure
 used to obtain the photometry for the original frames. This process is repeated several times in order to test a large number of
 artificial stars (60,000 in the case of each of our SMC fields). The injected and recovered magnitudes of the artificial stars,
 together with the information from those that were lost, provide the necessary information about all the observational effects
  present in the photometry.

\begin{figure}
\plotone{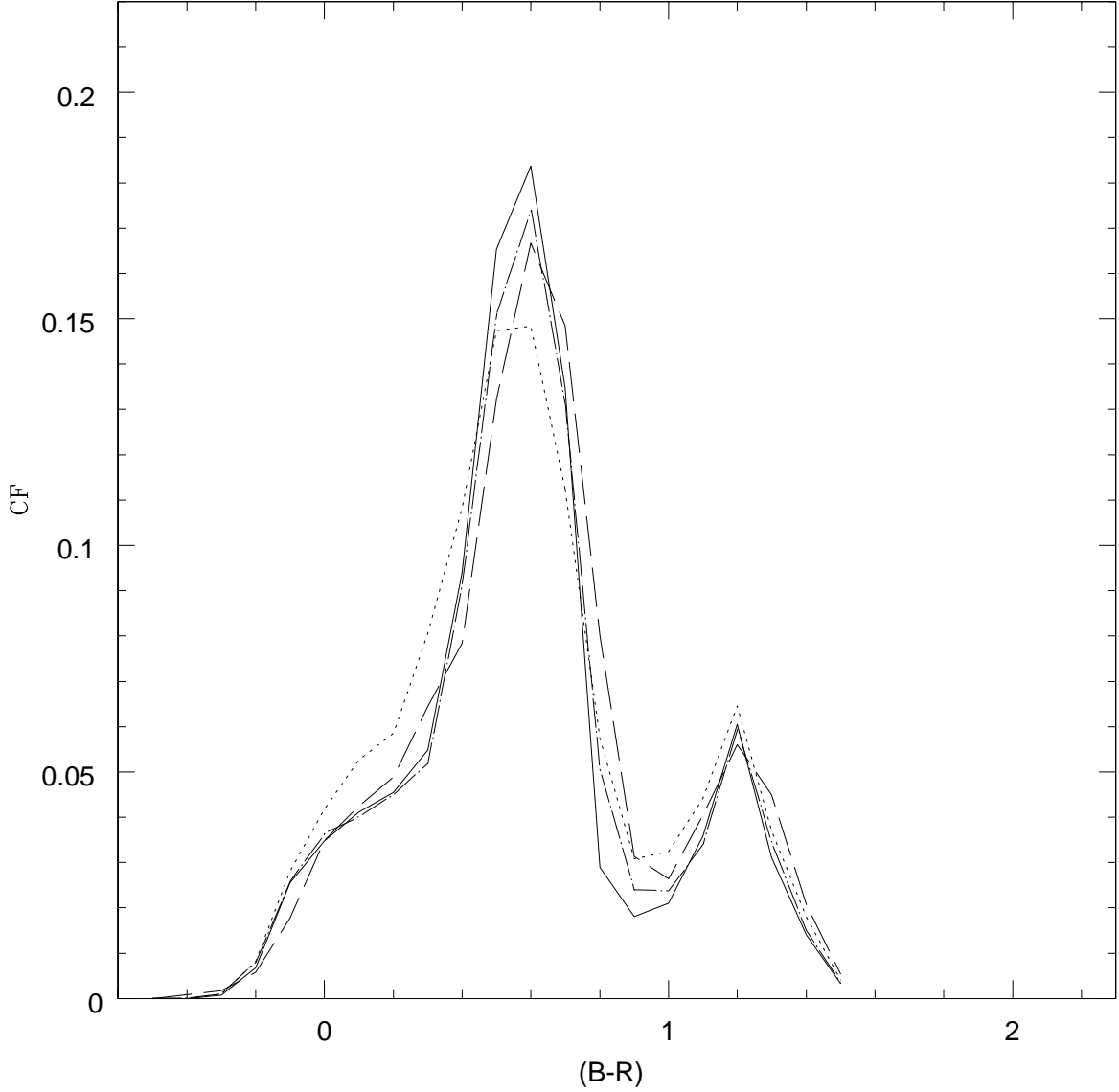}
\caption{Color functions of a synthetic CMD integrated over the magnitude 
range -1$\leqslant$M$_{R}$$\leqslant$3.5 assuming different levels of crowding in the data.
 The solid line represents the CF
of the original synthetic CMD, the dotted line corresponds to the CF of the synthetic CMD, in which the observational errors of  
 field smc0057 have been simulated, the dashed line is the CF of the synthetic CMD with the
simulation of errors from field qj0111, and dot-long dashed line is the CF of the synthetic CMD with 
errors in field smc0049 simulated.} 
\label{dispersadas}
\end{figure}
 
 We will discuss here how the observational errors affect the CF of a few representative fields.
  Those located closer to the SMC center (e.g. smc0057 and qj0111) are more affected by crowding 
   than the rest of the fields. To quantify these observational errors we have simulated them for
    different SMC fields in the synthetic CMD of Figure~\ref{docfsynt}. 
    Following Gallart et al. (1999), we will call the latter ``model CMDs''.
  The model CFs (corresponding to the model CMDs) corresponding to fields
 smc0057, qj0111, and smc0049 are shown in Figure~\ref{dispersadas}, together with the original CF (the CF obtained from the 
  synthetic CMD).
 The observational errors may produce significant variations in the appearance of the CF of our more central, crowded field smc0057.
 Simulating the observational
 errors of this field in the synthetic CMD, we obtain a model CF (dotted line)
 which differs substantially from the CF of the original synthetic CMD: the absolute maximum in its CF is smaller than in the CF of the
 synthetic CMD, and the blue elevation is bigger because faint stars on the low MS
  are lost much more frequently than the brighter blue stars that constitute the young MS.
  The dispersion introduced in the CMD when the observational errors are simulated produces an absolute maximum that is slightly wider
   as shown in the dotted line
  in Figure~\ref{dispersadas}.
 This shows how crowding
 could simulate a non-constant SFH, mimicking a recent enhancement of the star formation, similar to the one
 shown in blue in panel (b) of Figure~\ref{docfsynt}.
 Although still present, the effects of crowding are less severe in the case of field qj0111 since the CF of the corresponding model
CMD (dashed line) resembles quite closely
 the CF for the synthetic CMD. In this field, confusion is mainly affecting the {\it yellow} stars so the main maximum is
shorter than in the synthetic CF.
 In the case of smc0049, it can be seen that the CF of the
 model CMD (dot-long dashed line)
 is similar to the one of the synthetic CMD (solid line), indicating that crowding is affecting
  very little the CMDs of this part of the galaxy (at $\sim$2.9$\degr$).
In conclusion, only the CF of field smc0057 is substantially affected by observational errors.
The effects of crowding on field qj0111 can be considered as an upper limit for
fields qj0112 and qj0116. For fields
smc0100, qj0047, qj0102, smc0053, qj0037, qj0036, and qj0033 the crowding level is similar to that of smc0049. Therefore, we can safely
discuss the stellar content of 
all of our fields (except smc0057) by comparing their CFs straight with that of synthetic CMDs.

\begin{figure}
\plotone{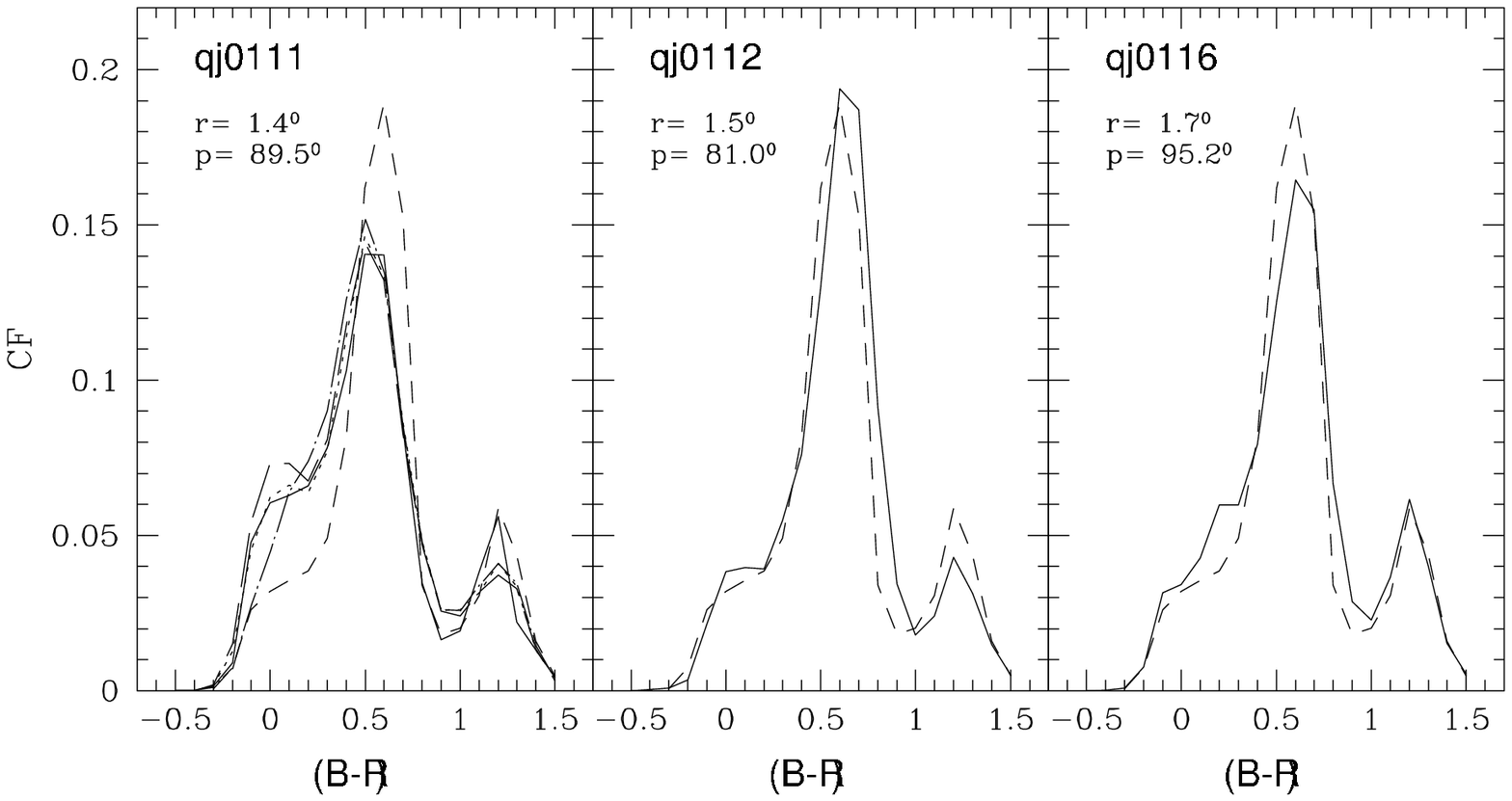}
\caption{CF for the observed CMDs corresponding to the Eastern SMC fields (solid lines), overlapped with the synthetic CF
(dashed lines) from panel (a) in Figure~\ref{docfsynt}, normalized to the number of stars in the 
region between -0.5$<$(B-R)$<$1.5 and -1$<$M$_{R}$$<$3.5.
In the first panel, corresponding to field qj0111, the CFs of synthetic populations with burst of star formation of 30\% and 50\% from 1
Gyr ago to now (dotted and long-dashed line respectively) and of 50\% from 2 Gyr ago to now (dot-long dashed line) were also
 overplotted, showing that there was a recent enhancement of the star formation around 1 Gyr ago in this field.}
\label{three_east_synt}
\end{figure}
\begin{figure}
\plotone{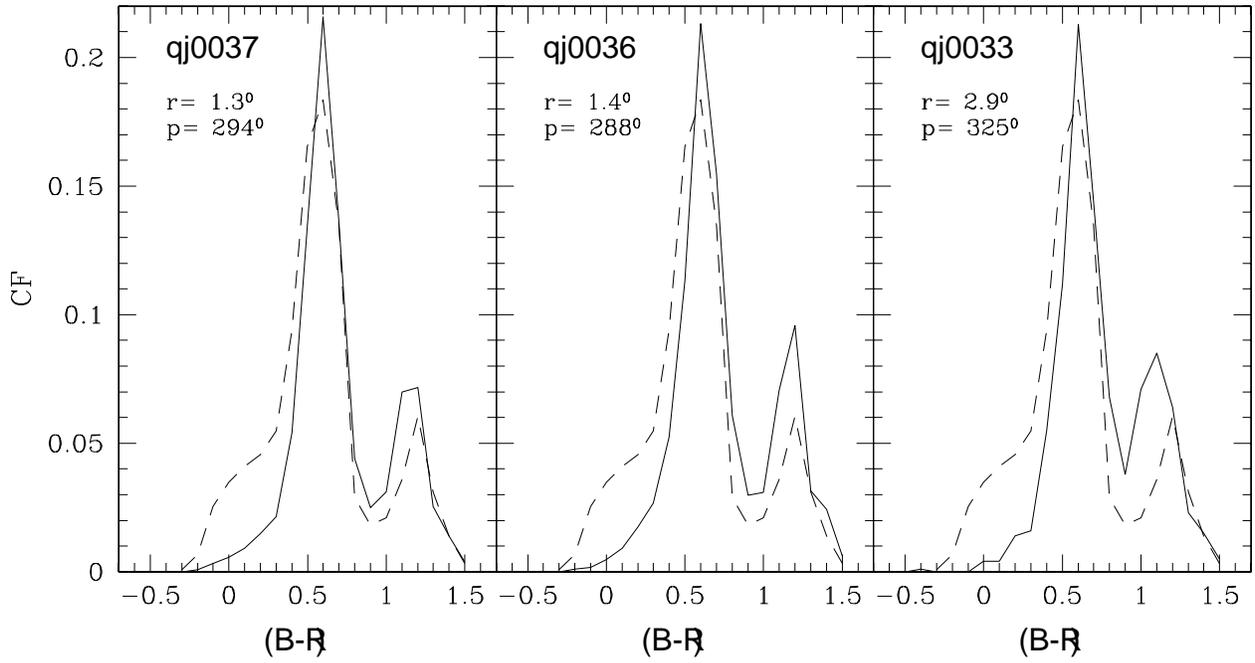}
\caption{CF for the observed CMDs corresponding to the Western SMC fields (solid lines), overlapped with the synthetic CF
(dashed lines) from panel (a) in Figure~\ref{docfsynt}, normalized to the number of stars in this
 region between -0.5$<$(B-R)$<$1.5 and -1$<$M$_{R}$$<$3.5.}
\label{three_west_synt}
\end{figure}
\begin{figure}
\plotone{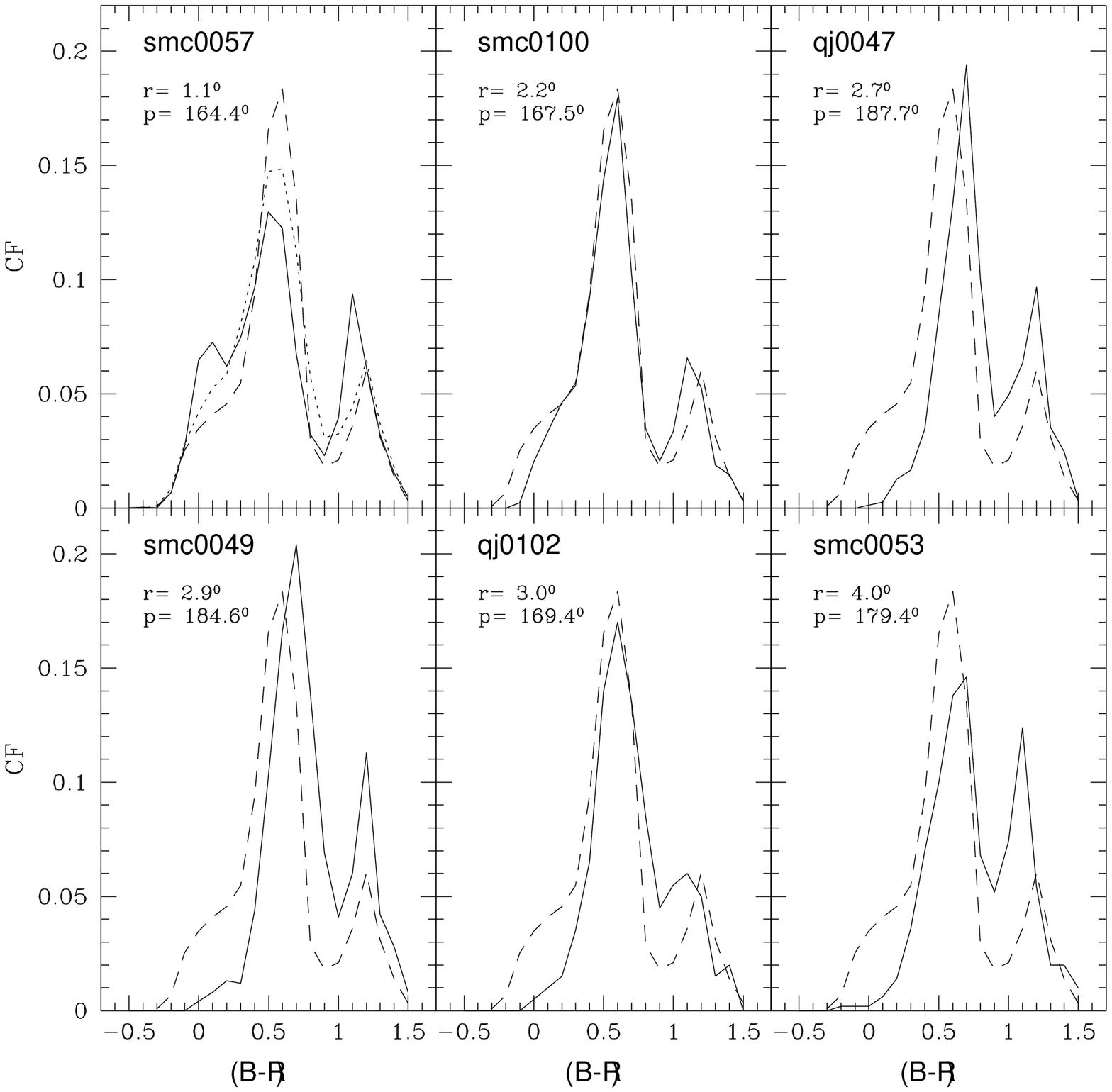}
\caption{CF for the observed CMDs corresponding to the Southern SMC fields (solid lines), overlapped with the synthetic CF
(dashed lines) from panel (a) in Figure~\ref{docfsynt}, normalized to the number of stars in this 
region between -0.5$<$(B-R)$<$1.5 and -1$<$M$_{R}$$<$3.5.
In field smc0057, was also superposed
 the CF of a model CMD (dotted lines) in which observational errors from this field were simulated (same as in Figure~\ref{dispersadas}).}
\label{six_south_synt}
\end{figure}

   Let's concentrate now on the observed CFs of each of our fields.
 In Figures~\ref{three_east_synt} to ~\ref{six_south_synt} the CF corresponding to the synthetic CMD (dashed lines)
 from panel (a) in Figure~\ref{docfsynt} and
the CFs (solid lines) from the observed CMDs were overlapped.
 Note that the 
 shape of the synthetic CF depends on the selection of metallicities and stellar evolution libraries. 
 However, small changes in the metallicity law, which result in synthetic CMDs with a color distribution compatible with the data have
 also very similar CFs.
Comparison between the observed and synthetic CFs computed
 assuming constant SFR(t) lets us to qualitatively analyze the SFH of each field.

Figure~\ref{three_east_synt} shows the CFs of the Eastern SMC fields.
In the first panel, corresponding to field qj0111, 
 synthetic CFs of synthetic populations
  with bursts of 50\% 1 Gyr ago (long dashed line), of 30\% 1 Gyr ago (dotted line), and of 50\% 2 Gyr ago
  (dot-long dashed line) were also overplotted.  
The CF of  field qj0111 shows some differences with respect to the CF of a population with constant SFR(t).
In particular, the observed CF has higher values than the synthetic one in the blue elevation
 (-0.3$\lesssim$(B-R)$\lesssim$0.2; age$\lesssim$1.1 Gyr) which indicates
  a SFH with enhanced star formation at young ages.
  This fact is also shown by comparing with the CF of synthetic 
  populations with bursts. In fact, the blue maximum and the main peak 
  of the CF with a burst of 30\% 1 Gyr ago and the ones of the observed CF match very well, pointing out that there was
   an episode around 1
  Gyr ago of enhanced star formation in this field which may have continued to the present time.  
 Though a comprehensive analysis of the CMDs will be presented in a forthcoming paper, 
 this superposition allows us to say that the observed CF is compatible with a SFH with 
  an increase in the star formation of $\sim$30\% in the last 1 Gyr.
   Also note that 
   crowding effects are not very significative in this field, as seen from the artificial star tests.  
 The CF of field qj0112 is a remarkable case where the CF corresponds almost exactly with that of a population with constant
  SFR(t) (at least in average). 
  The CF of qj0116
  presents a somewhat greater values of the CF in 
0$\lesssim$(B-R)$\lesssim$0.3 as compared with the CF of a CMD with constant SFR(t), 
 which may correspond to stars formed between $\sim$1-2 Gyr ago, possibly
indicating some enhancement in
the SFR(t) at those ages in comparison with more recent ages.
 Note that these three fields have a very similar CMD, and that simple comparison with isochrones gives
  the same information for all three:
 presence of stars of all ages.
 The differences in their CFs are a good illustration of the complementary information that the CF can provide.
  The presence in the observed CFs of all the features that characterize the synthetic CF is the counterpart,
 in this representation of the data, of the fact
that the CMDs are well populated over all age ranges. The additional information brought forward by the CFs
is the relative weight of the different age ranges in the population.

In Figure~\ref{three_west_synt}, the CFs of the Western SMC fields are shown. The almost complete
absence of the blue elevation in the
color interval -0.3$\lesssim$(B-R)$\lesssim$0.2, reflects the fact that there are few stars younger
than $\sim$1 Gyr. This is another example of the usefulness of the CFs: while
 in fields qj0036 and qj0037 there are stars until around
 the 0.1 Gyr isochrone, the CFs indicate that the relative 
 amount of stars between 0 Gyr and 2 Gyr is substantially less than in the case of
 a field with constant SFR(t).
The central maximum is slightly shifted to the red, possibly indicating that the SFR(t) started to decline around 3 Gyr ago.
 Note that this shift is unlikely due to an overall shift of the  observed CMD with
respect to the synthetic one, since the shift does not affect the red relative
maximum of the CF. In fact, this maximum is slightly shifted to the blue and higher in the CF of the observed 
CMD as compared with that of the synthetic one.  This is possibly due to field stars, which contribute more in relative numbers
 to the poorly populated CMDs. Actually, the observed red maximum is higher with respect to the theoretical one in the less populated
 qj0033.
 The composition of the population in these three Western CMDs seems to be very similar, even in the outermost field, qj0033. 

Finally, Figure~\ref{six_south_synt} shows the CFs of the Southern SMC fields. 
 In the case of field smc0057, the relative amount of young stars might be overestimated due to crowding effects
 if these were not taken into account. The overplot of the
 observed CF (solid line) with the synthetic (dashed line) and the model (i.e. after taking into account crowding effects, dotted line) 
 ones shown in Figure~\ref{six_south_synt} indicate that part of the enhancement of the blue 
 elevation in field smc0057 is
 produced by observational errors but that there is still 
 a true recent enhancement (0$\leqslant$t$\leqslant$3 Gyr) in the star formation.
The CF of field smc0100 is still almost compatible with a constant SFR(t) on average, with possibly a truncation of the SFR(t) at recent
ages, as also  hinted by the comparison of its CMD with isochrones in Figure~\ref{six_south_iso}.
 Moving further away from the SMC center, the elevations corresponding to the first structure
   in the CFs are less conspicuous, indicating that young stars are less common and
   intermediate-age stars dominate the fields.
The fact that the main peak is redder from field qj0047 on (except, maybe, in the case of field qj0102 which is
situated more towards the East than the other Southern fields),
 points out that there is a larger amount of old stars and, like in the Western fields, that possibly the SFR(t) started to decline around
 3 Gyr ago.
 As in the case of the Western field qj0033, the second,
  red relative maximum in the CFs gets bigger with increasing galactocentric distance. The reasons for this
are twofold. First, the amount of stars in the RC and in the RGB, relative to
 the total stellar content is larger at larger 
 distances from the center (even though not large differences among fields seem to be present beyong r$\thickapprox$2.7$\degr$),
  where the age of the stellar population is older on average. Second, 
 the contamination by foreground stars is larger relative to the total number
 of stars from the galaxy.  This last factor, may be the main player in the case of field smc0053. 
 
\section{SUMMARY} \label{summary}

We obtained {\it B} and {\it R}-band photometry of stars from 12 SMC fields observed during 4 different
campaigns with the C100 telescope at Las Campanas Observatory.
The spatial distribution of the
fields samples different parts of the galaxy, both in areas with large amounts of recent star formation such as the 
 ``Wing'' area and ``undisturbed'' regions at the   West and towards the South part of the galaxy, in which the stellar populations 
may be representative of the underlying population of the SMC formed prior to the events of star formation that shaped its current
irregular morphology.  
 The stellar content present in these SMC fields has been analyzed by means of a set of [{\it B-R, R}] CMDs that reach
 the oldest MS turnoffs (M$_{R}$$\sim$3.5) with an excellent photometric accuracy.

In this first aproach to study the SFH of our SMC fields, we found the following. 
 The fields in the ``Wing'' area contain very young blue stars and show 
  a broad MS turnoff/subgiant region and a wide range in luminosity of the RC, pointing out that star formation in these parts has
extended from at least $\sim$13 Gyr ago to the present with no substantial gaps,
as the areas around all the overlapped isochrones are well populated. 
Field qj0111 (the closest to the center of the SMC in this ``Wing'' area)
 seems to have experienced an enhancement in its recent
  star formation, compatible with an enhancement of 30\% in the SFR(t) over the last 1 Gyr.
   This is indicated by the height of the blue elevation its CF (color interval in 
  the range -0.3$\lesssim$(B-R)$\lesssim$0.2) as compared with that of CFs corresponding to synthetic populations with different SFR(t). 
   The
   other two fields analyzed in this direction seem to have had a SFR(t) close to constant on average.

Very little star formation has been going on from $\sim$1 Gyr ago until now in the Western fields, 
as indicated by the
 much lower amount of young blue stars and the lack of the blue elevation in the CFs.
 Considering the RC (formed by stars older than $\sim$1 Gyr)
  of fields qj0036 and qj0111, both located at the same galactocentric distance, we found that the number of stars is comparable, 
  indicating that the integrated SFR(t) from old ages and until 
  $\sim$1 Gyr ago in both fields is similar, and that there was a mechanism in the ``Wing'' area that
  triggered the star formation around 1 Gyr ago, in particular, in the location of field qj0111. 
  Field qj0033, though located further away from the center in the Northwest, presents a 
   distribution of stellar ages similar to that of 
   the two other Western fields, as seen from the resemblance of the three CFs and from its CMD, in which 
the area below the turnoff of the 4 Gyr isochrone is still quite populated.
In fact, the apparent differences between the CMD of field qj0033 and the other two are due to the much larger number of stars populating
those. If only a subset of stars of those CMDs are represented, to equal number of stars in the CMD of qj0033, the distribution of stars
around the different isochrones is very similar.
In these three fields, the bulk of stars seems to be older than 3-4 Gyr. A similar conclusion was reached by Dolphin et al. (2001)
for a field located
  in the Northwestern part and relatively close to field qj0033 (see Figure~\ref{SMCHI}),
   where they found a greatly decreased star formation in the past 2 Gyr.

 Regarding the fields observed toward the South, the young population is less important from r=2.7$\degr$ on.
The blue elevation in the CFs of 
the Southern fields is only found in the two fields closer to the center. In the case of smc0057, its shape is consistent with 
some enhancement of the recent SFH as compared to a constant SFR(t). 
From the CFs it is evident that 
field smc0100 has many more stars younger than 2 Gyr (relative to the total population
of the field) than the Western fields qj0036 and qj0037, even when it is almost 1$\degr$
further from the SMC center than the latters. Note that this field is located at the border of a more densely populated 
isopleth  than qj0036 and qj0037 and in the part in
which the trumpet-like distribution of the young population {\it fans out}.
Beyond  
$\sim$2.7$\degr$, there are only few stars younger than 3 Gyr old  
 and the differences between the fields are small (except for the fact that the density of stars in field smc0049 is larger than in
 the other two). 
 Finally, except for Eastermost field qj0102, the further South
 from the SMC center we go, the more is the main peak of the CF shifted to the red. This, together with the above facts,
 may indicate that the {\it main} epoch of star formation beyond $\sim$2.2$\degr$ in this direction ended around 3-4 Gyr ago.

\section{CONCLUSIONS} \label{conclu}
   
   The SMC shows a clear morphologic dichotomy between its young and old
population. The youngest component has an asymmetric distribution
which {\it fans out} towards the ``Wing'' area in the Northeast,
similarly to the HI distribution (Stanimirovi\'c et al. 1999), while
the RGB and AGB stars have a more symmetric, spheroidal
distribution. Using deep (down to $M_{R}=5.5$) CMDs in 12 small fields
that strategically sample different SMC regions, we are trying to
shed some new light into our knowledge of the SMC evolution, and
answer some of the questions posed in the Introduction.

We found that the underlying spheroidally distributed
population is mainly composed by intermediate-age and old stars, and its distribution does
not show strong galactocentric gradients. Our fields situated toward
the Northwest of the bar, at galactocentric distances from 1.3$\degr$
to 2.9$\degr$ contain very few stars younger than $\simeq$ 3 Gyr.
The comparison of their CFs with that of a synthetic population with constant SFR(t) indicates there has been an actual drop of the SFR(t)
at around this age until the present. Even 
though, some residual star formation seems to have continued up to now.
 No strong differences can be noticed between the CMDs
and CFs of the fields at 1.3$\degr$ and 1.4$\degr$ and the one at
2.9$\degr$. The apparent differences between these CMDs are mainly due
to the different number of stars in them, rather to a fundamentally
different composition of the stellar populations. Something similar
occurs in the case of the Southern fields beyond $\simeq 2.2
\degr$. In addition, we estimated that the SFR(t) at intermediate to old ages in fields located at the same distance from the SMC center
and at either side (e.g. qj0111 and qj0036) is similar.
 With the current analysis of the data we cannot provide a detailed SFH 
for these fields, and therefore we cannot confirm nor rule out periods
of quiescence as found in other studies that used shallower data. This
will be the subject of a forthcoming paper.

The three fields situated towards the East, in the ``Wing''
region, show very active current star formation. In the case of the field closer
to the center, qj0111, we estimated that the current star formation rate (from around 1 Gyr ago to the present) 
 in this field is $\sim$30\% larger than the SFR(t) averaged over the
galaxy's life. However, the other two northeastern fields, situated
just slightly further away from the center, have a CF which is
compatible with that of a constant SFR(t) over the whole galaxy's
history. Therefore, only the parts of the Wing closer to the center,
seem to be undergoing an exceptionally intense episode of star
formation at the present time.

At the radius of the innermost field ($\sim$1 kpc), we derived a crossing time 
of $\sim$10$^8$ yr, obtained adopting a measured velocity dispersion of 36 km/s (Carrera 2006, using the 
stars below the RGB tip corresponding to our SMC fields).
In the case of our outermost field ($\sim$4 kpc) the crossing time is 
$\sim$2$\times$10$^8$ yr.
 Large scale asymmetries should disappear within a few crossing times, and the population older than $\sim$1 Gyr is
 then expected to be well mixed at a radius of $\sim$1 kpc, while the population older than $\sim$2 Gyr is well mixed at $\sim$4 kpc.
 Hence, although our fields only cover little isolated parts of the SMC (as compared with the huge area embraced by HZ04),
  they are representative of the SMC populations older than $\sim$2 Gyr.

The larger amount of young stars present in the Eastern fields may be
related to a recent interaction between the SMC, the LMC, and the
Milky Way as is suggested by numerical models (e.g. Yoshizawa \&
Noguchi 2003). Note that the presence of a considerable young
population in the eastern fields and lack thereof in the western ones
is in good correspondence with the existence or absence of large
amounts of HI at the corresponding locations (Stanimirovi\'c et
al. 1999).  It is interesting to notice that this trend is not
maintained in the closest Southern field, smc0057, in which there is a
considerable amount of young population but the amount of HI is small.

The presence of a substantial amount of intermediate-age population in
our outermost SMC fields, could be a sign that a few gigayears ago HI gas
extended as far as $\sim$4$\degr$ and has since gradually
receded to the more central regions. 
Another possible explanation for this is that tidal interactions happened periodically, 
and then these stars dynamically mixed, resulting in old stars (those from previous passages) being 
symmetrically distributed and the younger ones only in the current tidal feature.
None of the studied fields is
dominated by an old stellar component, as one would expect for an old
stellar halo similar to the one of the Milky Way. This may mean that a
disk population is still dominating over a possible old halo, if it
does exist in the SMC.

\acknowledgments
We want to thank Antonio Aparicio and Santi Cassisi
 for many very useful opinions and discussions and
a careful reading of the manuscript. We are very grateful
to the anonymous referee for a critical and useful report that helped to improve the paper.
The authors acknowledge support by the Plan Nacional de Investigaci\'on Cient\'ifica, Desarrollo, e Investigaci\'on Tecnol\'ogica,
(AYA2004-06343). E. C. and R. A. M. acknowledge support by the Fondo Nacional de Investigaci\'on Cient\'{\i}fica y Tecnol\'ogica (No.
1050718, Fondecyt) and by the Chilean Centro de Astrof\'{\i}sica FONDAP (No. 15010003). This project has made generous use of the 10\%
Chilean time, and continuous support from the CNTAC and Las Campanas staff is greatly appreciated.

 			   
\end{document}